\documentclass[10pt]{article} 

\usepackage{setspace}

\usepackage{amsmath}
\usepackage{amssymb}
\usepackage{graphicx}

\usepackage{cite}

\usepackage{color} 
\usepackage{graphics,epsfig,amsfonts,amsmath,amssymb}
\usepackage{placeins}
\usepackage{epsfig}
\usepackage[labelfont=bf,labelsep=period,justification=raggedright]{caption}

\makeatletter
\renewcommand{\@biblabel}[1]{\quad#1.}
\makeatother

\date{}

\begin{document}

\begin{flushleft}
{\Large
\textbf{Bystander effects and their implications for clinical radiation therapy: Insights from multiscale in silico experiments}
}

\vskip.25cm
Gibin G Powathil$^{1}$, 
Alastair J Munro$^{2}$,
Mark AJ Chaplain$^{1\ast}$
Maciej Swat$^{3}$

\vskip.5cm
$^{\bf{1}}$Division of Mathematics, University of Dundee, Dundee, UK\\
$^{\bf{2}}$Radiation Oncology, Division of Cancer Research, University of Dundee, Ninewells Hospital and Medical School, Dundee, UK\\
$^{\bf{3}}$The Biocomplexity Institute and Department of Physics, Indiana University Bloomington, Bloomington, Indiana, USA\\
\vskip.5cm

$^{\ast}$Corresponding authors\\
Mark AJ Chaplain\\
Division of Mathematics\\
University of Dundee\\
Dundee, UK DD1 4HN\\
E-mails: chaplain@maths.dundee.ac.uk, gibin@maths.dundee.ac.uk

\vskip.5cm
\textit{Running title}: Modelling of radiation-induced bystander effects
\vskip.5cm
\textit{Keywords}: Multiscale mathematical model/ radiation therapy, radiation-induced bystander effects/ cell-cycle
\vskip.5cm
\textit{Conflict of Interest Statement}: The authors disclose no potential conflicts of interest.
\vskip.5cm
\end{flushleft}
\clearpage
\section*{Abstract}
Radiotherapy is a commonly used treatment for cancer and is usually given in varying doses. At low radiation doses relatively few cells die as a direct response to radiation but secondary radiation effects such as DNA mutation or bystander effects affect many cells. Consequently it is at low radiation levels where an understanding of bystander effects is essential in designing novel therapies with superior clinical outcomes. In this article, we use a hybrid multiscale mathematical model to study the direct effects of radiation as well as radiation-induced bystander effects on both tumour cells and normal cells. We show that bystander responses play a major role in mediating radiation damage to cells at low-doses of radiotherapy, doing more damage than that due to direct radiation. The survival curves derived from our computational simulations showed an area of hyper-radiosensitivity at low-doses that are not obtained using a traditional radiobiological model.

\section*{Introduction}

Radiotherapy is used in the treatment of 50\% of patients with cancer. The classic view of the action of ionising radiation is that it inactivates cells by causing the DNA damage which leads to cell death \cite{Prise2005}. However, depending upon circumstances, a greater or lesser proportion of the DNA damage may be repaired, and so the consequences, at the level of the individual cell, can vary from damage with complete repair, through damage with incomplete or inaccurate repair, to lethal damage \cite{Prise2009}. Cells vary in their intrinsic radiosensitivity \cite{Steel1991} and other factors also influence the cellular response to radiation: the oxygen level in the environment; the phase of the cell cycle; the repair capacity of individual cells. At the tissue level, the response will depend not just upon these cellular factors, but also on the ability of the cells that are critical for maintenance to repopulate the organ or tissue. The doses and fractionation schemes used in clinical radiotherapy represent a compromise between the desire to eliminate as many cancer cells as possible and the need to minimise the damage to normal cells and tissues.

Advances in molecular biology have expanded this classical view and it is now realised that signals produced by irradiated cells can influence the behaviour of non-irradiated cells - a range of phenomena known as  the ``bystander effect'' \cite{Blyth2011, Prise2009, Mothersill2004, Morgan2003a, Morgan2003b}. New technologies such as intensity-modulated radiotherapy (IMRT) allow irregularly shaped target volumes to be irradiated to high-dose whilst minimising the dose to vulnerable normal structures immediately adjacent to the tumour. The penalty paid however is an increase in the volume of normal tissue that is treated to a low-dose of radiotherapy \cite{Hall2003}. Since direct cell-kill is relatively low at low-radiation doses, bystander effects play a major role in determining the fate of cells and may be particularly relevant to radiation-induced carcinogenesis. Therefore, it is important to understand how novel therapeutic techniques might influence the occurrence and clinical consequences of bystander effects [13,17]. This is not an easy problem to investigate: it may be years before the consequences are expressed; the bystander effects may be swamped by effects on directly irradiated cells; the mediators and targets for  bystander effects are poorly understood \cite{Prise2009, Munro2009}. Mathematical and computational models offer the potential, at least in part, to circumvent these difficulties. By providing mechanistic insights into bystander phenomena these approaches will help to identify the key factors that are involved. Traditionally the linear quadratic model has been used as a useful tool for assessing radiotherapeutic treatments \cite{Powathil2007, Powathil2012, Powathil2013, Thames1982}. Furthermore, several mathematical models have been proposed to incorporate and study the effects of bystander phenomena \cite{Brenner2001, Little2004, Nikjoo2002, Nikjoo2003, Little2005, Shuryak2007, Richard2009}. Since the effects of radiation on tissue can manifest themselves in many ways at the cell, tissue and organ levels, we need systems-based multiscale model to better understand the impact of bystander signals on clinical outcomes. Multiscale approaches have the ability to incorporate several critical interactions that occur on different spatio-temporal scales to study how they affect a particular cell's radiation sensitivity, whilst simultaneously analysing the effects of radiation at the larger (tissue) scale \cite{Powathil2013, Richard2009,Ribba2006}. 

In this article, we develop the hybrid multiscale mathematical and computational model to study multiple effects of radiation and radiation-induced bystander effects on a growing tumour within host tissue. We use the new multiscale model to predict the effects of bystander signals on tissue treated with different radiation dosage protocols and analyse its implications for radiation protection, radiotherapy and diagnostic radiology.

\section*{Materials and Methods}

The multiscale mathematical model is developed by incorporating intracellular cell-cycle dynamics, an external oxygen concentration field and various effects of irradiation, including radiation-induced bystander effects that occur at multiple spatial and temporal scales. A comprehensive overview of the multiscale model with equations and parameter values can be found in previous papers by Powathil et al.  \cite{Powathil2012b, Powathil2013, Powathil2013b} and in the Supplementary Materials. Here, we investigate the effects of varying doses of ionising radiation on single cells and how this affects their non-irradiated neighbours (``bystander phenomena''). We use a hybrid multi-scale modelling approach to simulate the growth and progression of the tumour cells incorporating intracellular cell-cycle dynamics and microenvironmental changes in oxygenation status \cite{Powathil2012b, Powathil2013}. The intracellular cell-cycle dynamics are modelled using a set of five ordinary differential equations that govern the dynamics of key proteins involved in cell-cycle regulation \cite{Powathil2013, Novak2003, Novak2004}. The changes in the oxygen concentration within the domain of interest are incorporated into the model using a partial differential equation, governing its production, supply and diffusion \cite{Powathil2012, Powathil2013}. The details of these can be found in the Supplementary Materials as well as previous articles by Powathil et al. \cite{Powathil2012b, Powathil2013}. 

The simulations start with a single initial tumour cell in G1-phase of the cell-cycle at the center of a host normal tissue. Repeated divisions of the tumour cells - governed by the system of ordinary differential equations (ODEs) modelling the cell-cycle dynamics Ð produce a cluster of tumour cells. We have assumed that the normal cells divide only when there is free space in the neighbourhood (following the ODEs modelling the cell-cycle dynamics but with longer cell-cycle time), thus avoiding uncontrolled growth. Following the Cellular Potts Model methodology \cite{Glazier2007} we measure time in units of Monte Carlo Steps (MCS). In our model a single MCS corresponds to 1 hour of real time. The diffusion constants for oxygen and bystander signals are $2.5\times 10^{-5}$ cm$^2$s$^{-1}$ \cite{Powathil2012b} and $2\times 10^{-6}$ cm$^2$s$^{-1}$ \cite{Ballarini2006, McMahon2013}. We used an explicit Forward Euler numerical integration method to solve the PDEs governing oxygen and bystander signal dynamics. To avoid numerical instabilities we used a smaller time step in the numerical solutions of the PDE which required multiple calls to diffusion subroutine in each MCS (1000 for oxygen and 100 for bystander signals). Here, the decay rate of the radiation signal is assumed to be $\eta_s=0.021$ m$^{-1}$\cite{McMahon2013} and the rate of production of the bystander signal is assumed to be  dose-independent free parameter, normalised to 1 within the nondimentionalised equation. The parameters representing the decay and the production rates may vary over a very wide biologically plausible range, depending on the molecular nature of the signal, cell type and extracellular environment. Although some of the {\it in vitro} data suggest an infinite propagation of signals, measurements of bystander effect propagation in three-dimensional tissue-like systems \cite{Belyakov2005} and {\it in vivo} studies involving partial-body irradiation of mice \cite{Koturbash2006} show a finite range \cite{Shuryak2007}. This large variability suggests that these parameters together with the bystander response threshold and the probabilities (Figure \ref{model2}) can be further adjusted to study and reproduce various data sets, depending on the information on several factors such as molecular nature of the signal, cell type and extracellular environment.
 
At each step, all the cells are checked for the concentrations of intracellular protein levels and their phases are updated. If [CycB] is greater than a specific threshold (0.1) a cell is considered to be in G2 phase and if it is lower than this value, a cell is in G1 phase \cite{Powathil2013, Novak2003, Novak2004}. If the [CycB] crosses this threshold from above, the cell undergoes cell division and divides (along randomly chosen cleavage plane).  As the cells proliferate, the oxygen demand increases and in some regions, the concentration of oxygen falls below a threshold value (10\% of oxygen), making the cells hypoxic. Hypoxic cells are further assumed to have a longer cell-cycle due to the cell-cycle inhibitory effect of p21 or p27 genes expressed through the activation of HIF-1 under hypoxia \cite{Hitomi1998, Goda2003, Pouyssegur2006}(incorporated into Eq (1) of ODE system in Supplementary Materials). Furthermore, if the oxygen level of a cell falls below 1\%, that cell is assumed to enter a resting phase with no active cell-cycle dynamics \cite{Goda2003}.

To study the radiation effects, and to compare the direct and indirect effects of radiation, cells are assumed to be exposed to varying doses of radiation for 5 days, once a day starting from time=500 hr. The radiation is considered to affect the targeted cells either by direct effects through the direct induction of DNA double-strand breaks or by indirect effects through the radiation-induced bystander effects \cite{Prise2009, Mothersill2004, Morgan2003a, Morgan2003b}. Figure \ref{model1} illustrates various radiobiological effects of cell irradiation with the light green area indicating the direct effects and grey area showing the territory of the bystander effects. The direct effects of irradiation are modelled using a modified cell-based linear quadratic model, incorporating the effects of varying cell-cycle and oxygen dependent radiation sensitivities and other intracellular responses \cite{Powathil2013}. The survival probability of a cell after radiation is traditionally calculated using a linear quadratic (LQ) model \cite{Sachs1997}. Following Powathil et al. \cite{Powathil2013}, the modified linear quadratic model to study a cell's response to the irradiation is given by:
\begin{align}
S(d)=\exp\left[\gamma \left(-\alpha.\text{OMF}.d-\beta (\text{OMF}.d)^2 \right) \right].
\label{SensitivityLQ}
\end{align}
The parameter $\gamma$ accounts for the varying sensitivity due to the changes in cell-cycle phase and varies from 0 to 1, depending on a cell's position in its cell-cycle phase. As studies indicate, G2 /M phase cells are assumed to show maximum sensitivity (1) to radiation, while cells in G1 phase and resting cells are assumed to have a relative sensitivity of 0.75 and 0.25 respectively. We further assumed that normal cells that are not in the proliferative phase are less responsive to the radiation with a sensitivity of 0.25. The parameters $\alpha$ and $\beta$ are called sensitivity parameters and are cell/tissue specific while $d$ represents the radiation dosage. The effects of varying oxygen levels on a cell's radiation response is incorporated into the modified LQ model using the oxygen modification factor (OMF) parameter given by:
\begin{align}
\text{OMF}=\frac{\text{OER}(pO_2)}{\text{OER}_m}=\frac{1}{\text{OER}_m}\frac{\text{OER}_m.pO_2(x)+K_m}{pO_2(x)+K_m}
\label{OMF}
\end{align}
where $pO2(x)$ is the oxygen concentration at position $x$, \text{OER} is the ratio of the radiation doses needed for the same cell kill under anoxic and oxic conditions, $\text{OER}_m=3$ is the maximum ratio and $K_m = 3$ mm Hg is the $pO2$ at half the increase from 1 to $\text{OER}_m$ \cite{Powathil2013, Titz2008}. The model also assumes that after a low dose exposure to irradiation ($<5$Gy), about 50\% of the DNA damage is likely to be repaired within a few hours, increasing the survival chances of the cells and hence the final survival probability is written as:
\begin{eqnarray}\label{LQ_repair}
S^*(d)=\left\{
\begin{array}{lcl}
S\qquad \qquad \qquad \qquad \qquad \quad ~ d>5\\
\\
S+(1-S)\times 0.5 \qquad \qquad d\le 5.\\
\end{array}
\right. 
\end{eqnarray}
Furthermore, in calculating the radio-responsiveness of the irradiated cells, we have considered the effects of dynamic changes in radiosensitivity occurring post-exposure due to the redistribution of cells within the cell-cycle, repopulation of the tumour cell mass, reoxygenation of the tumour, and DNA repair delay in calculating the radio responsiveness of the irradiated cells. A detailed description can be found in Powathil et al. \cite{Powathil2013} and Supplementary Materials.

Experimental evidence shows that in addition to the {\it direct damage} due to radiation, irradiated cells produce distress signals, to which all neighbouring cells (i.e. both irradiated and non-irradiated) respond \cite{Prise2009}. One of the {\textit {in vitro}} studies has shown that these signals produced by the irradiated cells reach a maximum after 30 minutes of radiation and remain steady for at least 6 hours after the radiation \cite{Hu2006}. They have also showed that the signals could be transferred anywhere within the experimental dish \cite{Hu2006}. The indirect radiation bystander effects are produced by these radiation-induced signals sent by the irradiated cells that are directly exposed to the radiation \cite{Prise2009}. To model the effects of radiation damage to individual cells and to account for bystander effects we consider a field of bystander signal concentration ($B_s(x,t)$) which by diffusing to nearby cells produces probabilistic responses to these bystander signals i.e. the single bystander signal concentration field serves as a proxy for the multiple real bystander signals that affect cells adjacent to radiated regions in real, live tissue. Motivated by the experimental results, we assume that the spatio-temporal evolution of the signals is modelled by a reaction-diffusion equation, incorporating the production and decay of the signals from the irradiated cells, given by:

\begin{align}
&\frac{\partial B_s(x,t) }{\partial t}=\underbrace{D_{s}\nabla^2 B_s(x,t)}_{Diffusion}+\underbrace{r_{s} \text{cell}_{\text{Rad}}(\Omega,t)}_{Production}-\underbrace{\eta_{s}B_s(x,t)}_{Decay}
\label{chemo_equation}
\end{align}
where $B_s(x,t)$ denotes the strength or concentration of the signal at position $x$ and at time $t$, $D_s$ is the diffusion coefficient of the signal (which is assumed to be constant), $r_s$ is the rate at which the signal is produced by an irradiated cell, $\text{cell}_{\text{Rad}}(\Omega,t)$ ($\text{cell}_{\text{Rad}}(\Omega,t)=1$ if position $x\in \Omega$ is occupied by a signal-producing irradiated cell at time $t$ and zero otherwise) and $\eta_s$ is the decay rate of the signal. The bystander cells will then respond to these signals in multiple ways with various probabilities when the signal concentration is higher than a certain assumed threshold. To study the radiation-induced damage to both tumour cells and normal cells, we have assumed that the tumour cells grow within a normal cell population.

The multiple biological responses of the responding neighbouring cells include cell death, mutation induction, genomic instability, DNA damage and repair delay \cite{Prise2009,Prise2005, Blyth2011}. These biological effects are illustrated in the Figure \ref{model1}. Figure \ref{model2} shows the schematic diagram of the various biological responses of radiation that are incorporated into the computational model. It is assumed that all cells that undergo cell-death due to the direct effect (calculated based on LQ survival probability) emit these signals while the surviving irradiated cells that are under repair delay produce these bystander signals with some probabilities ($\textrm{Pt}_s=0.5$ and $\textrm{Pn}_s=0.2$ with $\textrm{Pt}_s>\textrm{Pn}_s$ in Figure \ref{model2}). It has been observed experimentally that not all cells respond to radiation-induced bystander signals and in addition, tumour cells are much more sensitive to bystander signals than normal cells are \cite{Prise2005, Millan2012}. Hence, we assume that the tumour cells respond to the bystander signals with a higher probability than that of normal cells ($\textrm{Pt}_b=0.3$ and $\textrm{Pn}_b=0.2$ with $\textrm{Pt}_b>\textrm{Pn}_b$). The cells that respond to these radiation-induced bystander signals react in various ways as illustrated in Figure \ref{model1} and we consider some of these bystander responses within the multiscale model as shown in Figure \ref{model2}. Depending on the signal concentration, these responses can be either protective or damaging \cite{Prise2005,Shuryak2007}. Here, these threshold intensities are taken to be $Kn_1=Kt_1=3$ and $Kn_2=Kt_2=4$ (signal concentration is normalised with production rate), as shown in Figure \ref{gradient1}.B, assuming that cells require a higher intensity threshold to respond to the bystander signals and there is a continuous response at least until 6 hours \cite{Hu2006}. However, depending on the cell line specific data, one could increased or decrease this threshold accordingly.

Protective responses involve delay to repair the DNA damage caused by the signals; damaging responses involve radiation-induced bystander cell-kill and mutagenesis. Normal cells that are responding  to the bystander signals are assumed to undergo repair delay for upto 6 hours if the concentration of bystander signal is higher than a threshold ($Kn_1<B_s(x,t)<Kn_2$ with $Kn_1=3$ and $Kn_2>4$) \cite{Hu2006}. If the concentration is greater than this threshold ($B_s(x,t)>Kn_2$) then with a series of probabilities the cell will undergo bystander signal induced cell death (if random probability is less than $\textrm{Pn}_{k}=0.1$) or will mutate and initiate a radiation-induced cancer (if random probability is greater than $\textrm{Pn}_{k}=0.1$). Similarly, tumour cells responding to the bystander signals either undergo bystander signal induced cell death ($B_s(x,t)>Kt_2$) or repair delay ($Kt_1<B_s(x,t)<Kt_2$ with $Kt_1=3$ and $Kt_2>4$) depending on the signal concentration \cite{Prise2005, Prise2009}. As Mothersill and Seymour \cite{Mothersill2006} indicated, it is hard to determine the exact probabilities or thresholds by which these bystander responses of normal or tumour cells occur but we provide a sensitivity analysis of these probabilities on the cell-kill in the Supplementary Materials. A comprehensive list of rest of the parameters used in the model can be also found in the Supplementary Materials.

\section*{Results}
The clinical advantage of radiation therapy critically dependent on a compromise between the benefit due to the radiation-induced tumour cell-kill and the potential damage to normal tissues \cite{Munro2009}. In addition to the direct cell-kill, radiation can also cause multiple biological effects that are not directly related to the ionising events caused due to the irradiated dosage. Figure \ref{model1} illustrates various radiobiological effects of cell irradiation with the light green area indicating the direct effects and grey area showing the territory of the bystander effects and Figure \ref{model2} shows the relevant effects that are included in the present hybrid multiscale model. Firstly, we simulate a growing tumour within a cluster of normal cells using the multiscale mathematical model that is described in the methodology section. Figure \ref{gradient1}.A shows the spatio-temporal evolution of the host-tumour system at times=100, 300, 500 and 700 hrs. The colours of each cell represent their specific position in their cell-cycle and the hypoxic condition, as illustrated in the figure legend. The figure also shows the changing morphology of the growing tumour and the development of fingerlike projections at the tumour boundary as seen in most of the human malignancies. This also indicates that the host-tumour interaction can play a major role in spatial distribution and development of a growing tumour through competitive interactions \cite{Kosaka2012}. 

To study the effects of radiation, we consider three types of radiation delivery and exposure to treat a growing tumour within a cluster of normal cells. In most of the cases of current radiation delivery planning, the dosage received by both tumour volume and surrounding normal volume is not uniform. While the gross tumour volume (GTV) and clinical target volume (CTV) receive high dose radiation, rest of the surrounding area, the irradiated volume receive a decreasing dosage depending on their distance from the GTV or CTV. In case 1, we consider that a tumour is treated homogeneously to the prescribed dose per fraction and the rest of the irradiated normal tissue receives lower doses with decreasing intensity depending on their location from the tumour rim, as would occur in the clinical context. Case 2 analyses the radiation effects assuming that only tumour cells are fully exposed to the given radiation dose, sparing any normal cells which is an ideal scenario for radiation planning and in case 3, we consider the worst case scenario where both tumour and surrounding normal cells receive the same given dosage. In the following, we will discuss the results of these three cases and their clinical implications. 

In Figure \ref{gradient1}.B, we plot the spatio-temporal evolution of the host-tumour system before and after one of the five doses of irradiation at time=548 hr. Plots in the upper row of the Figure \ref{gradient1}.B show the spatial distribution of cancer and normal cells with bystander signal producing cells labelled in light blue. These plots show that after the irradiation, most of the signalling cells are located at the area exposed to relatively lower doses of radiation (time=552 hr) as higher doses of irradiation leads to direct cell-kill.  The plots also show that the radiation-induced cell-death creates empty spaces, reactivating the growth or normal cells in the neighbourhood (yellow and purple cells). The plots in the lower panel of Figure \ref{gradient1}.B show the change in the concentrations of radiation-induced bystander signals emitted by the cells that are labelled in light blue. The scaled values shown in the colormap indicate the strength of these bystander signals with maximum value 3. The signal concentrations beyond this value trigger bystander responses from either normal or tumour cells. The plots show that at time=546 hr, there are bystander signals of lower strength than those previously produced by signalling cells during the fraction at time=524 hr. Although the strength of these signals represented here are weak at time=548 hr, depending on the number of signalling cells and irradiation fractions, this could build up over a number of radiation fractions to reach a damaging level that might trigger bystander responses. After the irradiation at time=548 hr, the concentration of bystander signals increase and stay above the threshold value for around 6-10 hours, before dropping below the threshold as it is observed in the experimental studies \cite{Hu2006}. Moreover, there is a dynamically changing, heterogeneous concentration of bystander signals throughout the host-tumour system. The simplified use of threshold value could be justified by the experimental observation \cite{Hu2006} that the radiation-induced bystander effects are a distance-dependent phenomenon and that the responses in cells close to the signalling cells are significantly greater than those in more distant cells.

The direct and indirect effects of radiation when the host-tumour system is treated with specific doses of radiation for the three cases described above are shown in Figure \ref{gradient2}. The figure shows the total cell-kill due to the radiation and the contributions from the direct hit and other bystander responses for various dose per fractions. The plots show that, in all three cases, the total cell-kill increases with an increase in the dose per fraction. However, at the lower doses (dose = 0.25 Gy and dose = 0.5 Gy) the contributions from the radiation-induced bystander cell-kill predominate - as seen in previous experimental studies \cite{Prise2009, Hu2006}. Moreover, as opposed to the direct cell-kill calculated using LQ model, the bystander responses are similar even with an increase in the radiation dosage. In case 1 (Figure \ref{gradient2}. A), the surrounding normal cells that are exposed to a decreasing intensity of radiation dosage also respond to the radiation-induced bystander signals with bystander signal induced cell-kill or increased DNA damage contributing to carcinogenesis. However, in case 2 (Figure \ref{gradient2}. B), the radiation-induced bystander effects are minimal and does not, contrary to case 1, induce any mutation. The data from Figure \ref{gradient2}. C for the case 3 shows that there is a significant increase in the radiation-induced direct cell-kill, although the bystander cell-death is similar to that in the previous scenarios. The plots also show that the irradiation of normal cells may also lead to DNA mutation, increasing the chances of carcinogenesis.

Figure \ref{tumour} shows the survival fractions of the host-tumour system for all three cases when it is treated with 5 fractions of radiation with varying doses and is further qualitatively compared to experimental results by \cite{Joiner2001}. The survival fractions of the system are compared with and without the bystander signal induced cell-kill, assuming that the total number of cells for the control case is 1000. The plots show a region on high cell-kill at the doses 0.25 Gy and 0.5 Gy (the tumour is exposed to the maximum dose while normal cells receive less than the maximum dose) as compared with doses greater than 0.5 Gy. The region of hyper radio-sensitivity at low dose levels or inverse-dose effect, as seen in the Figures \ref{tumour}.A- \ref{tumour}.C is also observed in several experimental studies, as shown in Figure \ref{tumour}.D \cite{Prise2005, Joiner2001, Marples2000} and is not predicted using the traditional LQ models. The survival curves for case 3 are plotted in Figure \ref{tumour}.B which compared with the previous cases, shows less pronounced hyper radio-sensitivity. This is due to the increase in direct cell-kill of both normal and tumour cells since all the cells are exposed to the given maximum dosage. Additional results showing the multiple factors involved in the direct and indirect effects of radiation are given in Supplementary Materials.

\section*{Discussions}

Radiation-induced bystander effects have shown to play a major role in determining the overall effects of radiation, especially at low dose rates \cite{Prise2009, Prise2005, Blyth2011, Munro2009}. Although, the precise mechanisms underlying the induction and response of bystander signals are not yet fully understood, several molecular and intracellular cell communication processes have been widely implicated in mediating bystander effects \cite{Prise2009, Prise2005, Mothersill2004}. The cells that are in direct contact with each other are thought to support bystander signalling through the gap junctions \cite{Prise2009}. The bystander responses of the cells that are not in close contact are mediated through the release of diffusive protein-like molecules such as cytokines from the cells that are irradiated. Although, radiation-induced bystander effects have been extensively studied experimentally, their relevance and role in clinical radiation treatment and human carcinogenesis risk remain to be explored further \cite{Munro2009, Sowa2010}. Most of the experimental studies investigating the bystander responses are based on {\it in vitro} systems where cells are grown within media and showed no significant spatial effects as signals seem to diffuse rapidly throughout the medium \cite{Hu2006,Schettino2003}. However, spatial heterogeneity has been observed in more tissue like structures \cite{Belyakov2005} and hence consideration of spatial variation is important while studying the bystander effects in clinically relevant systems. The complex nature of various radiobiological interactions within a living organism after radiation exposure further limits detailed \textit{in vivo} and clinical investigations \cite{Munro2009, Blyth2011}. Nevertheless, our continued pursuit of bystander experimental studies using more complex \textit{in vitro} and tissue models, highlights the importance of identifying key processes and parameters that may play vital role in radiation-induced bystander responses. Here, we have presented a mathematical and computational modelling approach to study the direct and indirect effects of radiation and in particular, radiation-induced bystander effects, after exposing the host-tumour system to varying radiation doses. 

We considered the computational analysis of a growing tumour within a cluster of normal cells, incorporating those properties of individual cells (cell-cycle phase; external oxygen concentration) that influence the direct and indirect responses of cells to irradiation. The direct effects of radiation were studied using a modified linear quadratic model that incorporates some of the important factors responsible for radiation sensitivity such as cell-cycle phase-specific radiation sensitivity, improved survival due to DNA repair, and hypoxia. We have also considered the indirect effects of radiation though bystander effects, where the assumption is that irradiated cells produce bystander signals as a result of stress due to DNA damage. These signals diffuse within and around the irradiated volume. Computations involving bystander effects were carried out using probabilistic methods, assigning specific probabilities for the production of bystander signals and responses towards bystander signal concentration. The rest of the parameters in the model were either chosen from literature or extracted from experimental observations \cite{Hu2006, Powathil2013, Powathil2013b}. Here, we do not focus on explicitly fitting our model results to any particular experimental data which vary depending on multiple factors such as the nature of the experiment ({\it in vivo} or {\it in vitro} studies), the cell type or the molecular nature of the cell, but rather try to understand experimental observations and qualitatively study the effects of bystander effects on overall radiation effectiveness and responses. However, with the help of a relevant data set, if desired, our current model can be further tuned to reproduce various experimental results.

The computational models were then used to study the radiobiological effects of radiation considering three different total treatment volume and varying radiation doses per fraction. The results obtained from the model are qualitatively in good agreement with the experimental findings and clinical observations. In all three cases, the cell-kill due to the bystander effects dominated the total radiation cell-kill at low dose rates when the majority of the cells are exposed to low dose radiation, while the proportion of direct cell-kill increased with the increase in the radiation dosage (Figure \ref{gradient2}). However, the cell-kill due to the bystander effects remained relatively similar, irrespective of the varying doses per fraction. These findings are qualitatively in good agreement with the experimental findings by Hu et al. \cite{Hu2006}, who irradiated fibroblasts to study the spatio-temporal effects of bystander responses by calculating the fraction of DNA double strand breaks (DSBs). They found that within the irradiated area, the fraction of DSBs was high with higher doses (direct cell-kill) but the region outside the irradiated volume had lower but relatively similar rates of DSBs regardless of the dosage level (bystander effects). They also found that at lower dosage level, the fraction of bystander induced DSBs is almost equal to the fraction of DSBs (both bystander and radiation-induced) within the irradiated volume (Figure 3 in \cite{Hu2006}).

Figure \ref{gradient2} also shows the bystander responses of surrounding normal cells. Comparing cases 1, 2 and 3, it can be seen that when the both normal and tumour cells are exposed to the irradiation, the bystander responses of normal cells include bystander signal induced cell-death and DNA mutation potentially contributing to the carcinogenesis.  However in case 2, when the total treatment volume contains tumour cells only, no significant bystander responses are observed in normal tissue (except for repair delay associated with the DNA damage induced by bystander signals). This is in accord with the clinical observation that highly localised (small treatment volume) radiation is more effective than techniques using higher volumes, although this is still a matter of some debate \cite{Murray2013}.

The survival curves plotted for all three cases (Figures \ref{tumour}.A-\ref{tumour}.C) showed an inverse dose-effect: an increase in cell-killing at a range of low dose rates that would not be predicted by back-extrapolating the cell survival curve for high dose rates. These findings are qualitatively in consistent with the several experimental observations that showed a region of low dose hypersensitivity \cite{Prise2005, Joiner2001, Marples2000}. Figure \ref{tumour}.D shows once such experimental result where survival curves of asynchronous T98G human glioma cells irradiated with 240 kVp X-rays and measured using cell-sort protocol is plotted. In the Figure, the area of hypersensitivity can be observed with in the dose range of 0-1 Gy. Although, some of the experimental evidence suggests that increased DNA repair might contribute increased resistance at higher doses \cite{Joiner2001, Marples2000}, the present results suggest that radiation-induced bystander responses might contribute to this observed hypersensitivity \cite{Prise2005}. An increased number of actively signalling cells at lower doses could explain the dominance of bystander cell-death over radiation-induced cell death at low dose rates and thus contributing to an inverse-dose effect.

Analysing the direct and indirect effects of radiation for all three cases, it can be seen that the volume of exposure and the dose of radiation have major effects on total cell-kill and bystander responses. In case 1, more normal cells are killed when the tumour cells are irradiated with doses greater than 1.5 Gy, exposing the normal cells to doses from 0 Gy to 1.5 Gy. The greater cell-kill at higher doses reduces the number of bystander signal producing cells, resulting in lower bystander responses at higher doses. In short, at low dose rates, low direct cell-kill and moderate bystander cell-kill contribute to the total cell-kill while at higher doses, high direct cell-kill of tumour cells, moderate direct cell-kill of normal cells and low bystander cell-kill add to the total cell-kill. In case 2, the direct effects are based on the contribution from direct tumour cell-kill. At higher doses more tumour cells are killed, reducing the number of signalling tumour cells, while at low doses more tumour cells produce bystander signals, increasing the bystander response and cell-kill. As compared to cases 1 and 2, the survival curve for case 3, showed a less significant region of hypersensitivity at low doses. This is because when a uniform dose is given to the entire system, there is high cell-death of both tumour and normal cells due to direct irradiation and at low dose rates, all the cells are exposed to the given dose as oppose to the case 1 where they receive a range of doses from 0 to the maximum. In all three cases, the damages induced by the bystander effects on normal cells are minimal as we assumed that the normal cells are less likely to produce bystander signals, they have high repair capability (less direct effects) and they do not respond well to the surrounding bystander signals, as suggested by the experimental observations \cite{Prise2005, Millan2012}.

Our understanding of the role of radiation-induced bystander signals in mediating the risk of secondary cancers after treatment is limited. Most of the findings about the bystander effects are derived from \textit{in vitro} studies with artificial settings and limited clinical applicability. Multiscale mathematical models such as the one we present here can serve as powerful investigative tools, incorporating multi-layer complexities to understand and identify the multiple parameters that are significant in radiation-induced bystander responses. The computational model we have developed explores the spatio-temporal nature of radiation-induced bystander effects and their implications for radiation therapy. By explicitly incorporating a  consideration of bystander effects on tumours and normal tissues, our model can be used to enrich the information provided by traditional LQ models and thereby expand our knowledge of the biological effects of ionising radiation.

\section*{Acknowledgments} 
The authors gratefully acknowledge the support of the ERC Advanced Investigator Grant 227619, M5CGS: From Mutations to Metastases: Multiscale Mathematical Modelling of Cancer Growth and Spread. 


\clearpage
\section*{Figure Legends}
\begin{figure}[h!]
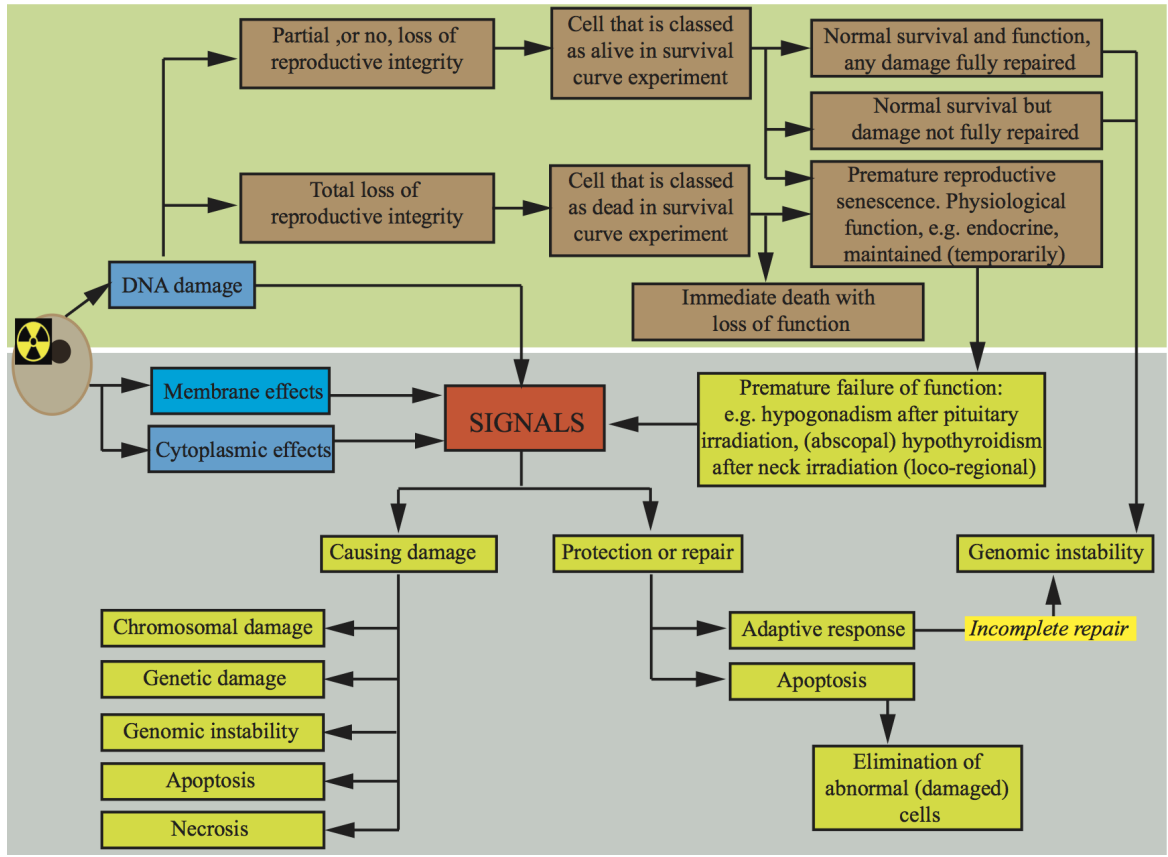

  \begin{center}
      \end{center}
  \caption{Diagram showing multiple biological effects of radiation. Here, classical radiation biology operates within the area shaded green and bystander effects operate within the area shaded grey.}
  \label{model1}
\end{figure}

\begin{figure}[h!]
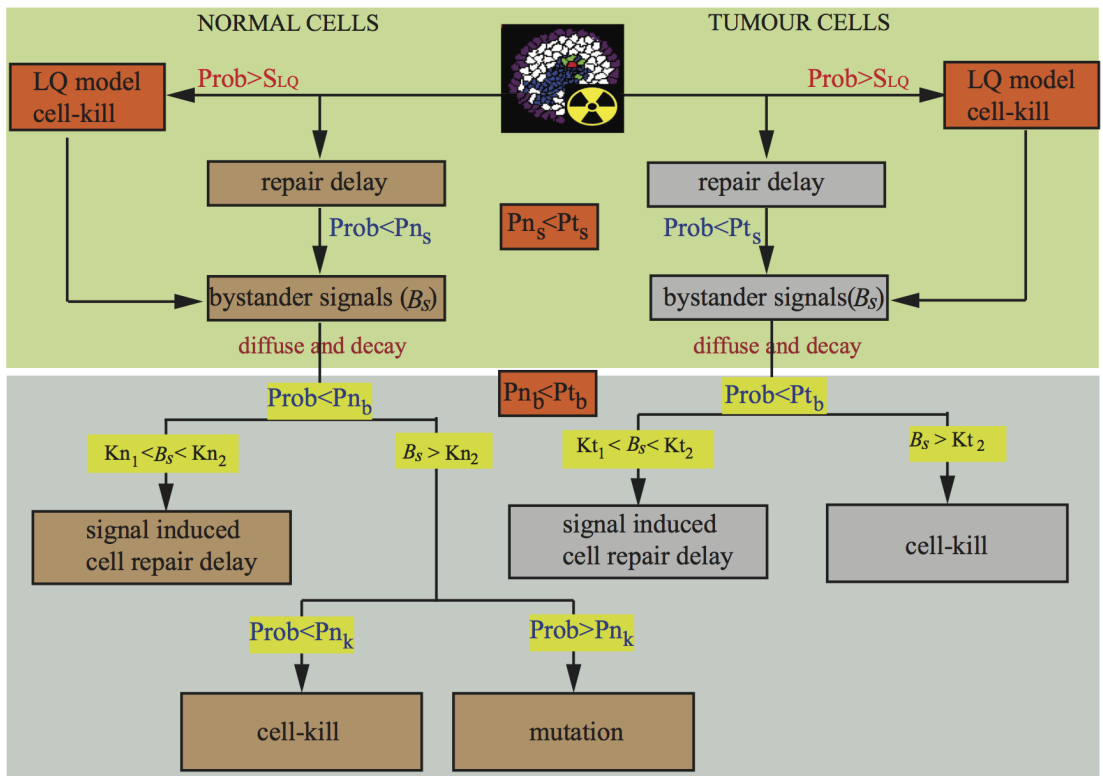

  \begin{center}
      \end{center}
  \caption{Diagram showing various interactions that are incorporated into the computational model when a growing tumour within normal tissue is irradiated. Here, we have added the responses of both normal and tumour cells. }
  \label{model2}
\end{figure}

\begin{figure}[h!]
  \begin{center}
      \end{center}
  \caption{Plots showing the spatio-temporal evolution of host-tumour dynamics with and without treatment. (A) Plots showing the growing tumour at four different simulation time steps and (B) Plots show the changes in the spatial distribution when the host-tumour system is irradiated at times=548. Upper panel shows the changes in cell distribution as well as the signalling cells after irradiation and the lower panel shows the distribution of bystander signals with colour map indicating various threshold values}
  \label{gradient1}
\end{figure}

\begin{figure}[h!]
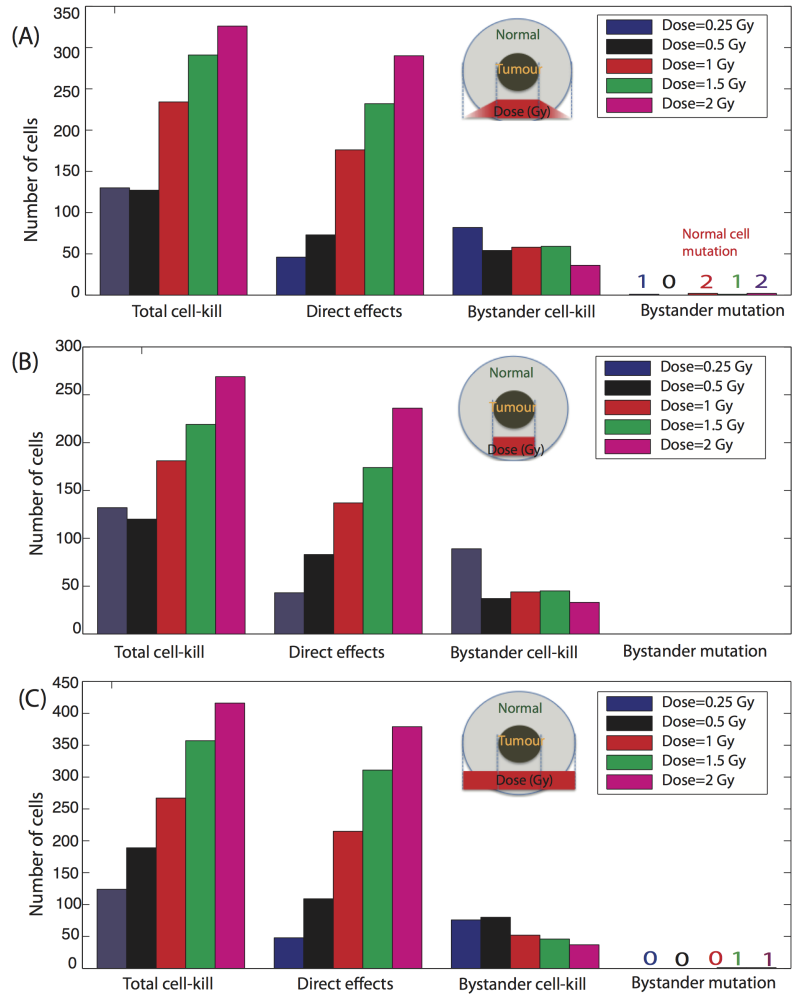

  \begin{center}
      \end{center}
  \caption{Plots show the number of cells killed under the direct effects and indirect effects of radiation and other bystander signal responses. (A) Case 1: tumour cells are exposed to full given radiation dose while the surrounding normal cells receive gradient of doses, (B) Case 2:  tumour cells are exposed to full given radiation dose while the surrounding normal cells are spared completely and (C) Case 3: tumour and normal cells are exposed to full given radiation dose.}
  \label{gradient2}
\end{figure}

\begin{figure}[h!]
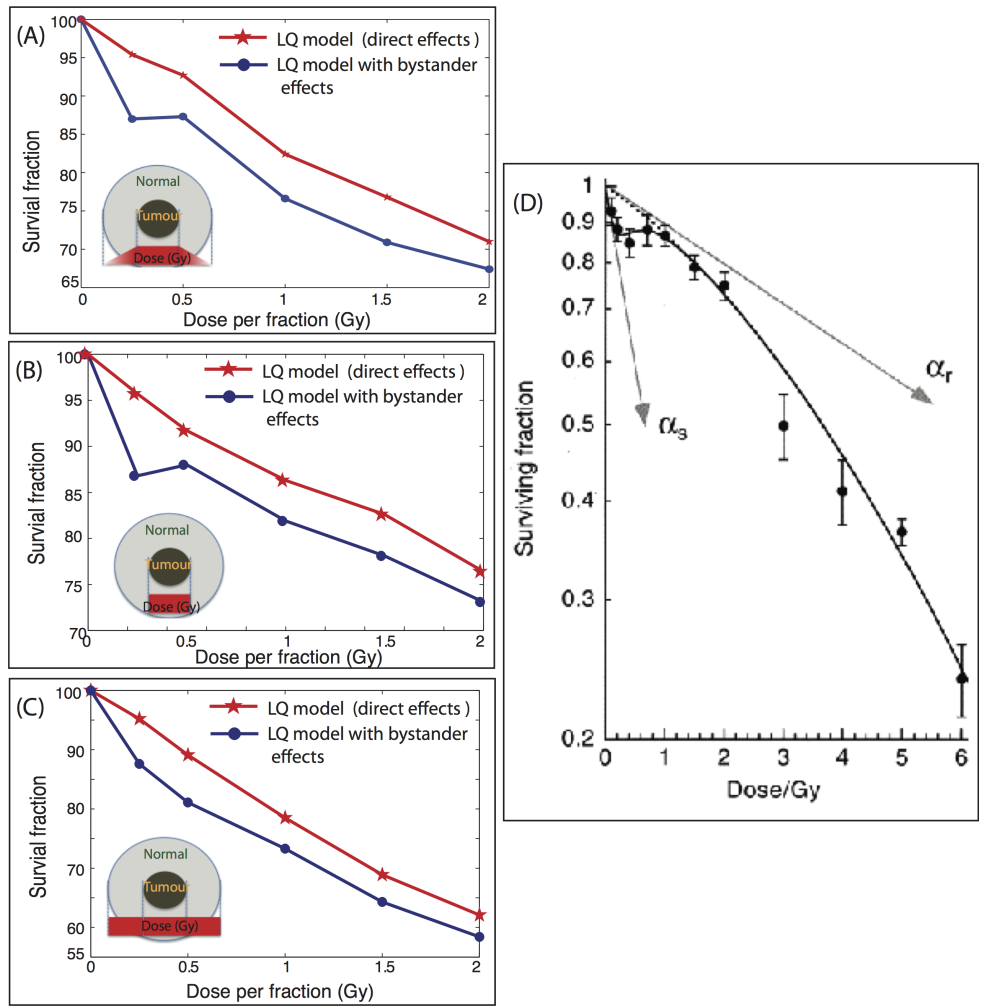

  \begin{center}
      \end{center}
  \caption{Plots show the differences in the survival fraction when bystander responses are considered. (A) Case 1: tumour cells are exposed to full given radiation dose while the surrounding normal cells receive gradient of doses, (B) Case 2:  tumour cells are exposed to full given radiation dose while the surrounding normal cells are spared completely (C) Case 3: tumour and normal cells are exposed to full given radiation dose. and (D) Experimental result: survival of asynchronous T98G human glioma cells irradiated with 240 kVp X-rays, measured using the cell-sort protocol (Figure from Joiner et al. \cite{Joiner2001}, used with copyright permission) }
  \label{tumour}
\end{figure}

\clearpage
\setcounter{figure}{0}
\section*{Figures}
\begin{figure}[h!]
  \begin{center}
      \includegraphics[scale=1.04]{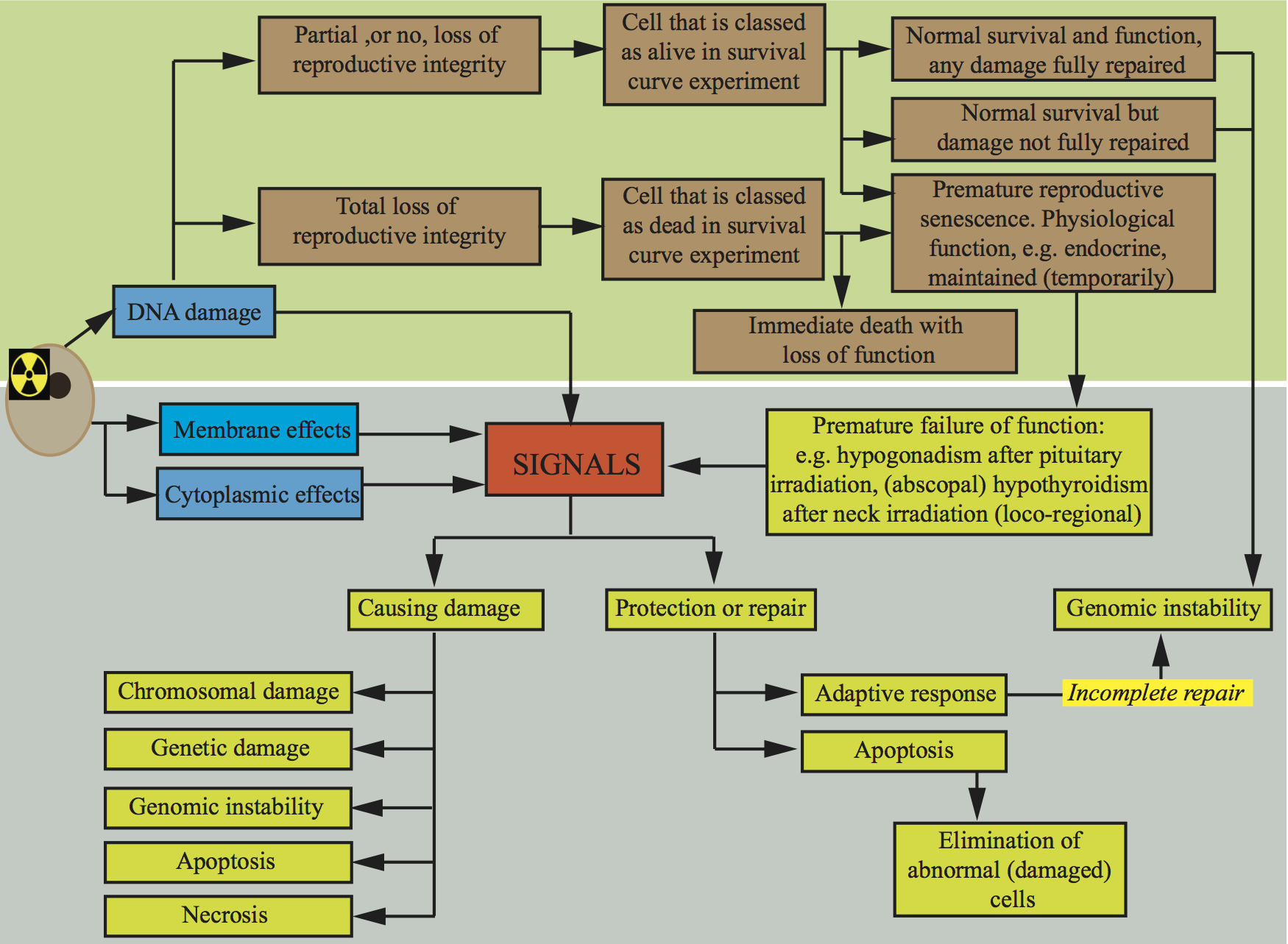}
      \end{center}
  \caption{Diagram showing multiple biological effects of radiation. Here, classical radiation biology operates within the area shaded green and bystander effects operate within the area shaded grey.}
  \label{model1}
\end{figure}

\begin{figure}[h!]
  \begin{center}
      \includegraphics[scale=1]{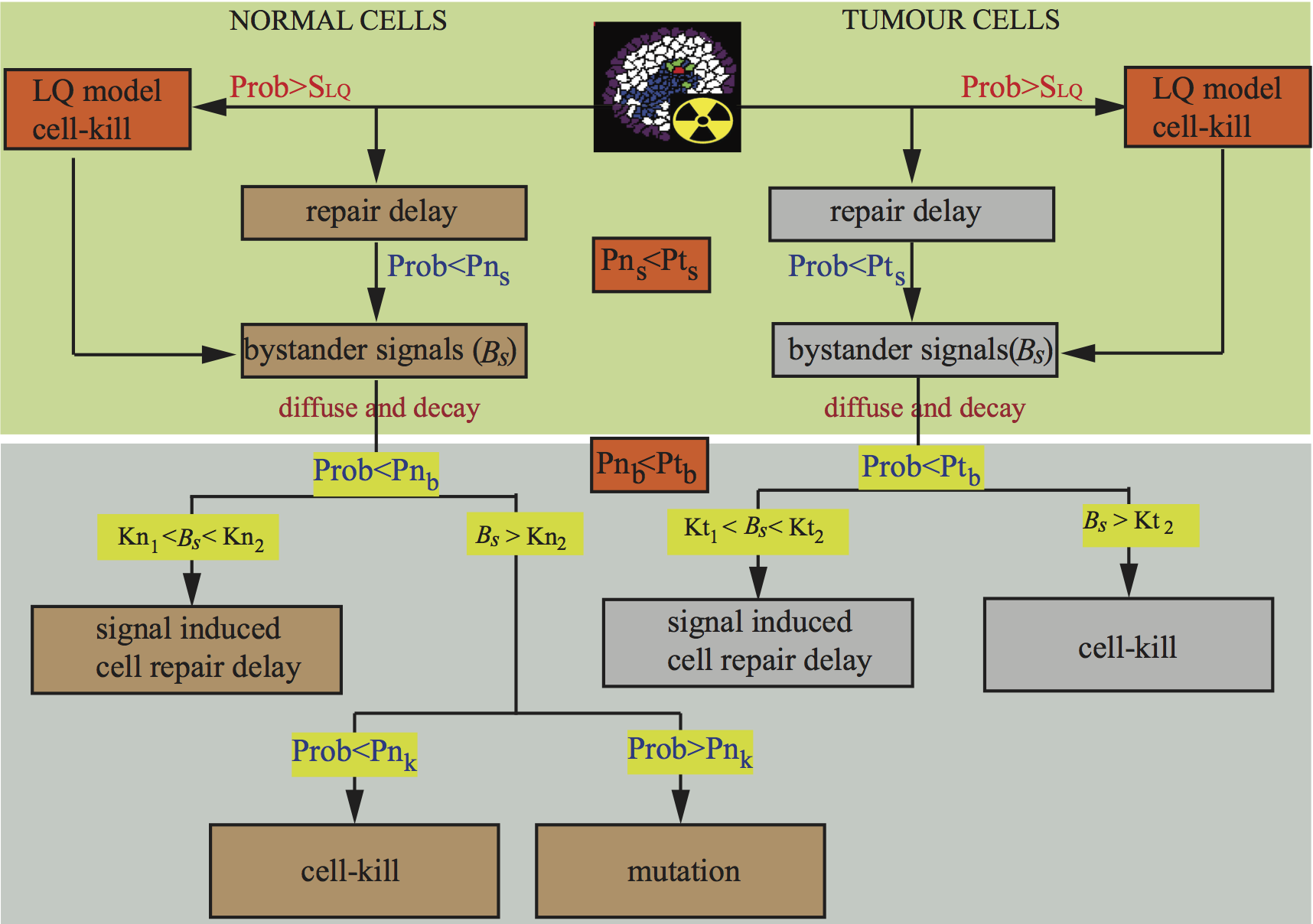}
      \end{center}
  \caption{Diagram showing various interactions that are incorporated into the computational model when a growing tumour within normal tissue is irradiated. Here, we have added the responses of both normal and tumour cells. }
  \label{model2}
\end{figure}

\begin{figure}[h!]
  \begin{center}
      \includegraphics[scale=.8]{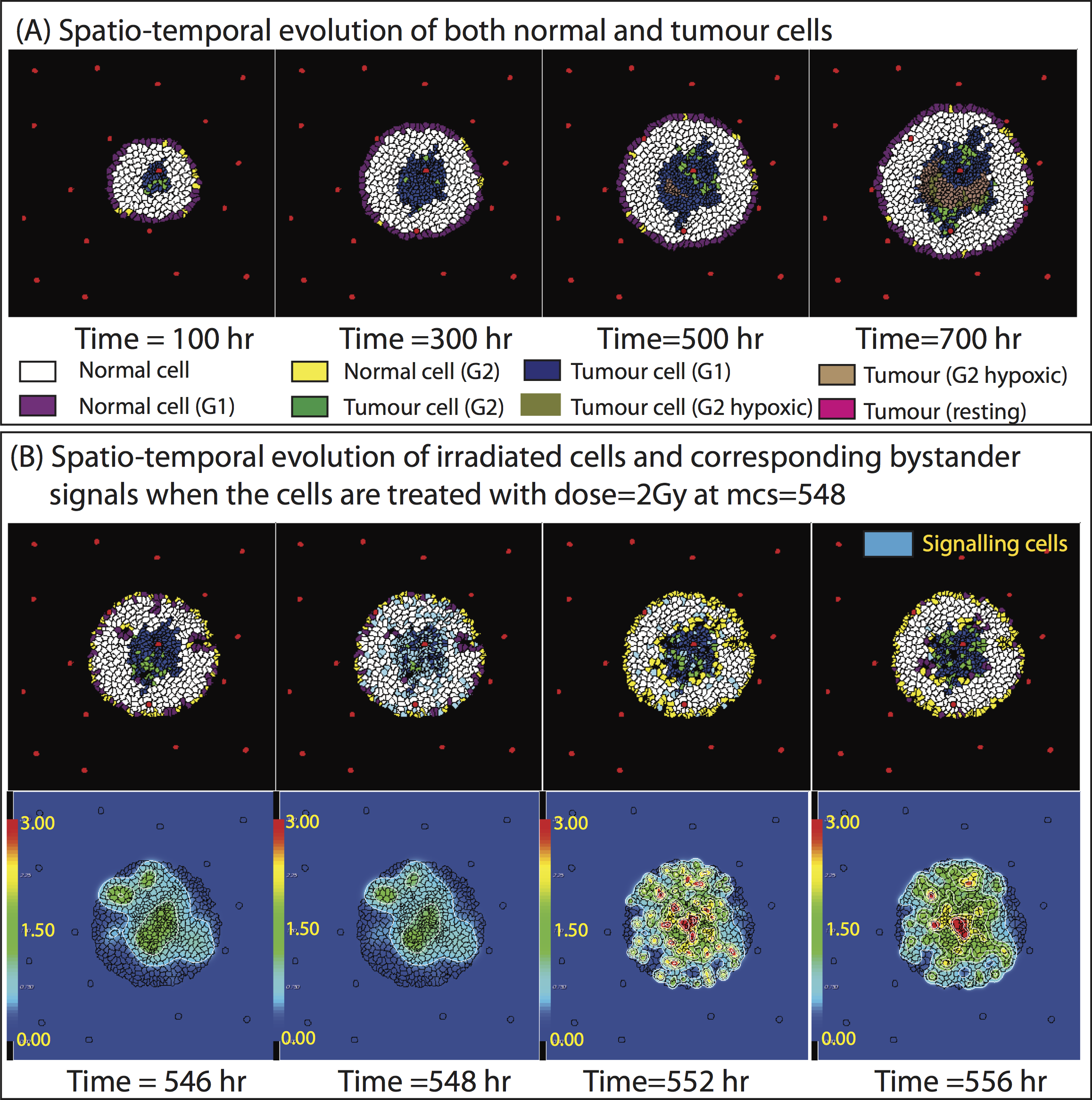}
      \end{center}
  \caption{Plots showing the spatio-temporal evolution of host-tumour dynamics with and without treatment. (A) Plots showing the growing tumour at four different simulation time steps and (B) Plots show the changes in the spatial distribution when the host-tumour system is irradiated at times=548 hr. Upper panel shows the changes in cell distribution as well as the signalling cells after irradiation and the lower panel shows the distribution of bystander signals with colour map indicating various threshold values}
  \label{gradient1}
\end{figure}

\begin{figure}[h!]
  \begin{center}
      \includegraphics[scale=.6]{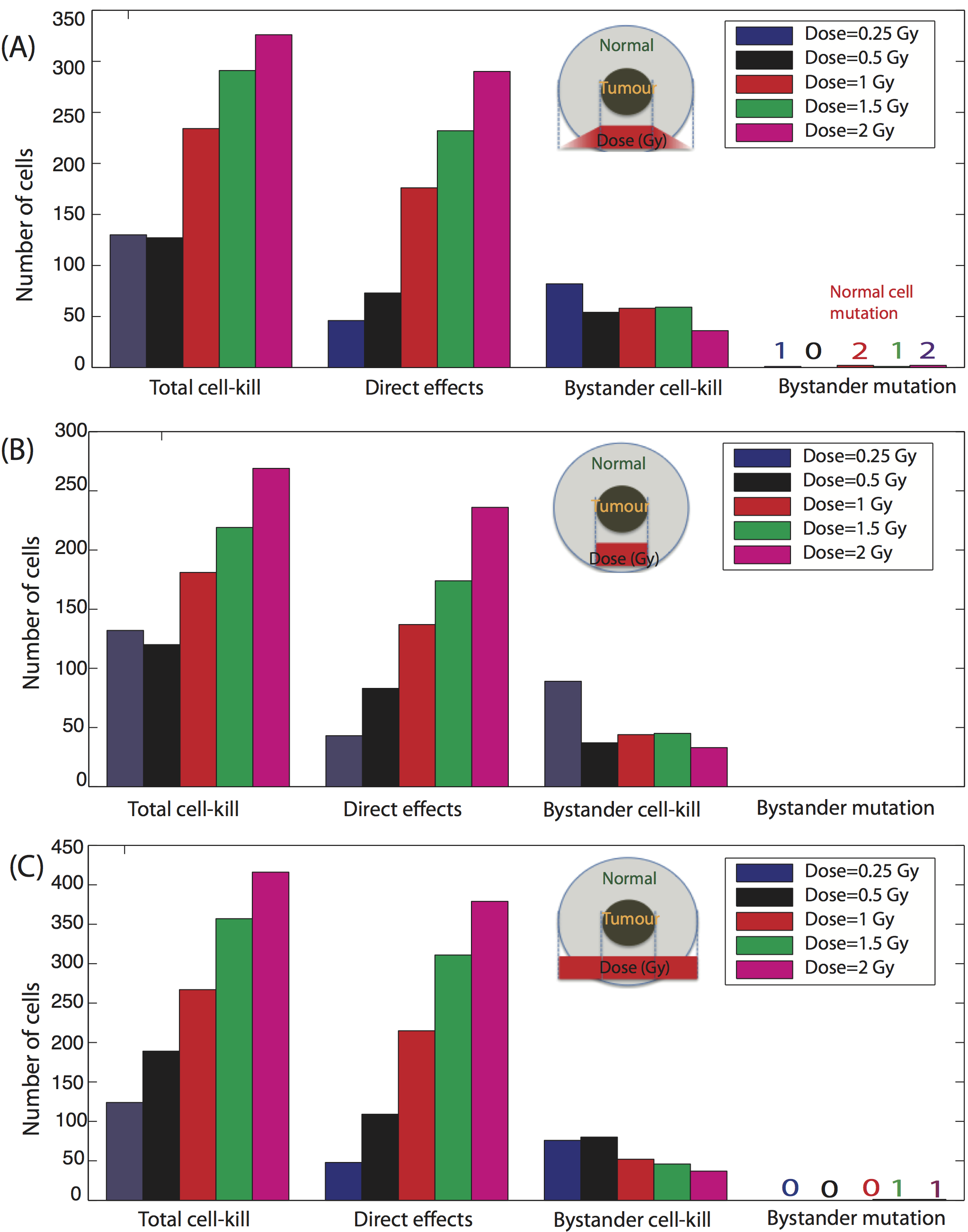}
      \end{center}
  \caption{Plots show the number of cells killed under the direct effects and indirect effects of radiation and other bystander signal responses. (A) Case 1: tumour cells are exposed to full given radiation dose while the surrounding normal cells receive gradient of doses, (B) Case 2:  tumour cells are exposed to full given radiation dose while the surrounding normal cells are spared completely and (C) Case 3: tumour and normal cells are exposed to full given radiation dose.}
  \label{gradient2}
\end{figure}

\begin{figure}[h!]
  \begin{center}
      \includegraphics[scale=.6]{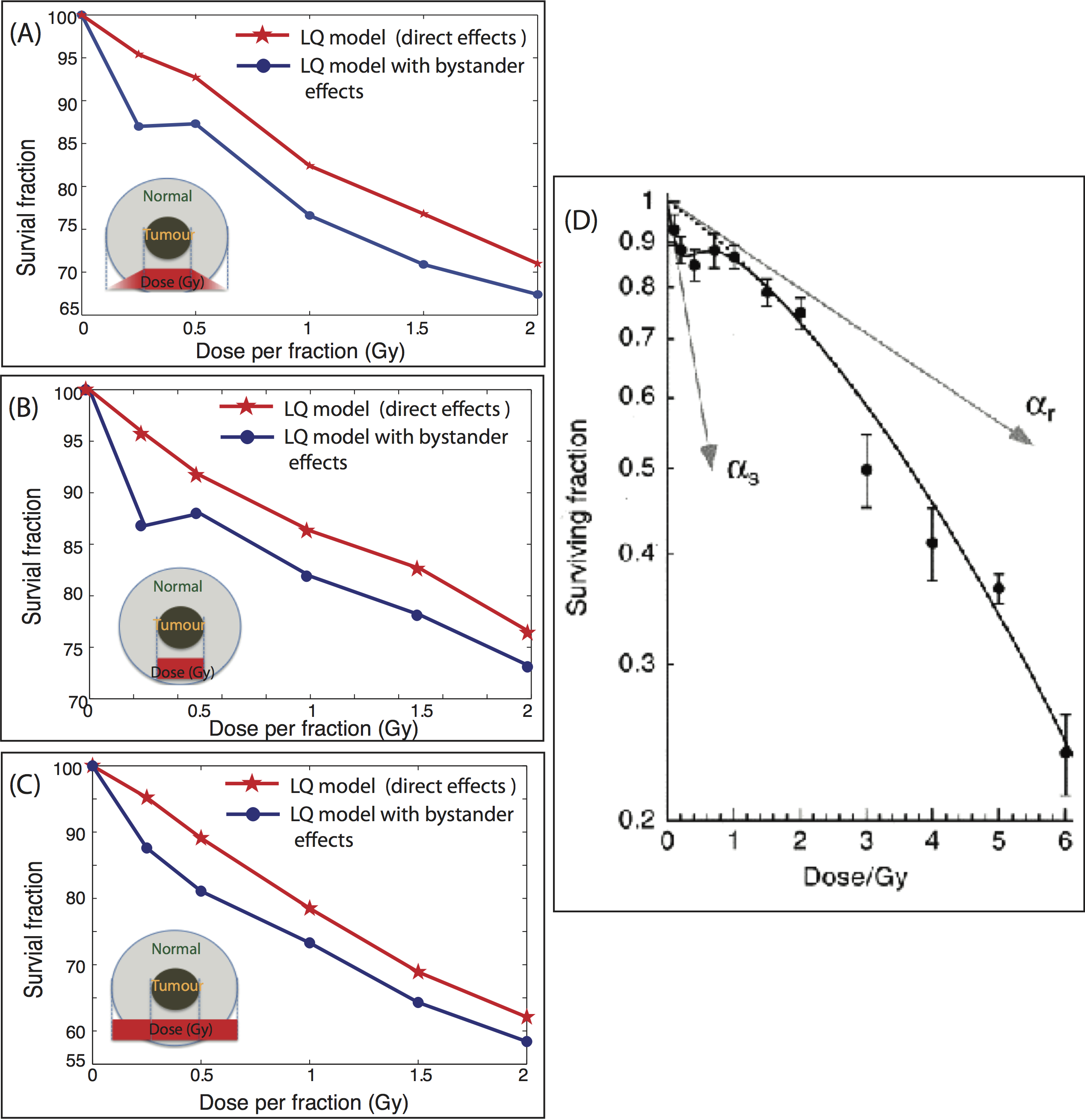}
      \end{center}
  \caption{Plots show the differences in the survival fraction when bystander responses are considered. (A) Case 1: tumour cells are exposed to full given radiation dose while the surrounding normal cells receive gradient of doses, (B) Case 2:  tumour cells are exposed to full given radiation dose while the surrounding normal cells are spared completely (C) Case 3: tumour and normal cells are exposed to full given radiation dose. and (D) Experimental result: survival of asynchronous T98G human glioma cells irradiated with 240 kVp X-rays, measured using the cell-sort protocol (Figure from Joiner et al. \cite{Joiner2001}, used with copyright permission) }
  \label{tumour}
\end{figure}

\end{document}


\renewcommand{\thefigure}{S\arabic{figure}}
\begin{center}
{\Large \textbf{Supplementary Materials}}
\end{center}

\vskip.5cm
\begin{flushleft}
{\Large
\textbf{Bystander effects and their implications for clinical radiation therapy: Insights from multiscale in silico experiments}
}
\\
Gibin G Powathil,
Alastair J Munro,
Mark AJ Chaplain,
Maciej Swat

\vskip1cm

\end{flushleft}


\begin{figure}[h!]
  \begin{center}
      \includegraphics[scale=.5]{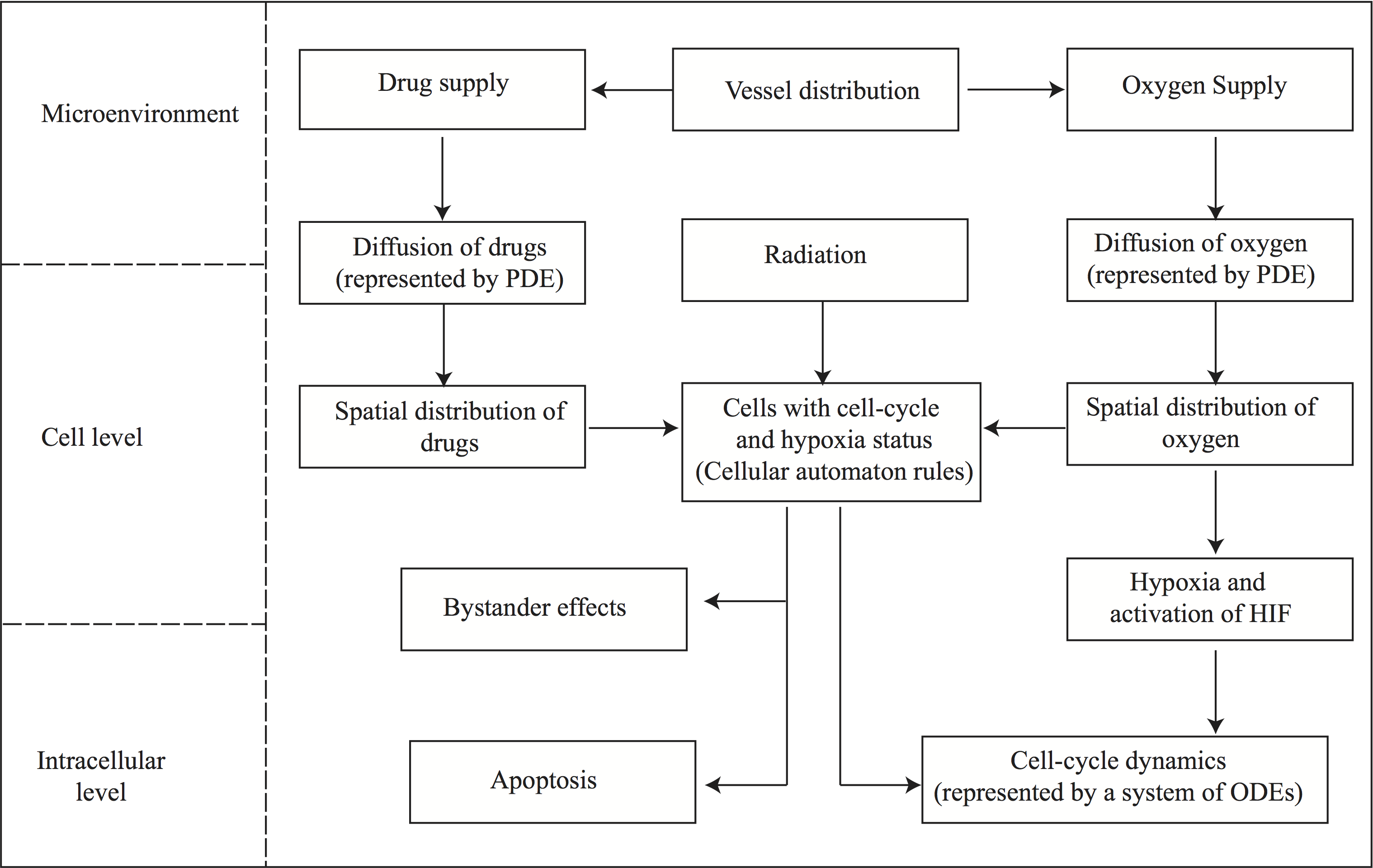}
      \end{center}
  \caption{Schematic diagram of the model showing the appropriate scales involved}
  \label{diagram}
\end{figure}
\subsection*{Cell growth: The CompuCell3D framework}

The computational simulations are carried out using the CompuCell3D framework developed by Glazier et al. (see http://www.compucell3d.org for full details). To simulate the growth of a tumour monolayer from a single cell we use a 2-dimensional lattice of size $300 \times 300$ pixels in the $x-$ and $y-$directions and it mainly consists of four major components. These are: (1) cells (either tumour or normal cells) whose spatio-temporal evolution is controlled by internal cell-cycle dynamics and the external microenvironment; (2) cross-sections of blood vessels from where the oxygen is supplied within the domain and (3) diffusive fields - oxygen concentration distribution and radiation-induced bystander signals.The initial configuration is one single cancer cell surrounded by a cluster of normal cells and a number of blood vessels. Division of the cells and hence growth of the tumour mass is driven by internal cell-cycle dynamics, modelled using a set of ordinary differential equations. These are incorporated into the Compucell3D framework at each Monte Carlo time step (MCS) of the simulation using Bionetsolver. Bionetsolver is a C++ library that permits easy definition of sophisticated models coupling reaction-kinetic equations described in SBML with the defined cells for execution in CompuCell3D \cite{Andasari2012, Powathil2013b}. Bionetsolver makes use of the SBML ODE Solver Library to implement reaction-kinetic network dynamics which can regulate the cell-cycle dynamics for each cell within the domain. The oxygen and drug concentrations are incorporated into the CompuCell3D as diffusive chemical fields modelled by PDEs governing the spatio-temporal dynamics. Regarding the time-scale of tumour growth, each MCS is assumed to be one hour in length and the parameters in the multiscale model are scaled accordingly. A schematic diagram of the model showing various scales involved is given in the Figure \ref{diagram}


\subsection*{Cell-cycle dynamics and intracellular heterogeneity}
Following Powathil et al. \cite{Powathil2012b,Powathil2013, Powathil2013b}, the growth and division of cells within this hybrid multiscale model is modelled using a system of ODEs that describes the changes in the concentrations of some of the key proteins that are involved in the cell-cycle mechanism. Here, we use a basic cell-cycle model adapted from Tyson and co-authors  \cite{Novak2001, Novak2003}. Using the kinetic relations they explained the transitions between two main states, G1 and S-G2-M, of the cell-cycle, which is (in their model) controlled by changes in cell mass. Here, the transition between phases is constrained by the cell's volume that increases after each MCS. The system of ODEs governing the cell-cycle dynamics is given by:

\begin{align}
&\frac{d\text{[CycB]}}{dt}=k_1-(k^{'}_2+k^{''}_2\text{[Cdh1]}+[\text{p27/p21}][HIF])\text{[CycB]},\\
&\frac{d\text{[Cdh1]}}{dt}=\frac{(k^{'}_3+k^{''}_3\text{ [p55cdc}_\text{A}])(1-\text{[Cdh1]})}{J_3+1-\text{[Cdh1]}}-\frac{k_4\text{[volume][CycB][Cdh1]}}{J_4+\text{[Cdh1]}},\\
&\frac{d \text{ [p55cdc}_\text{T}]}{dt}=k^{'}_5+k^{''}_5\frac{(\text{[CycB][volume]})^n}{J^n_5+(\text{[CycB][volume]})^n}-k_6 \text{ [p55cdc}_\text{T}],\\
&\frac{d \text{ [p55cdc}_\text{A}]}{dt}=\frac{k_7\text{[Plk1]}(\text{ [p55cdc}_\text{T}]-\text{ [p55cdc}_\text{A}])}{J_7+\text{ [p55cdc}_\text{T}]-\text{ [p55cdc}_\text{A}]}-\frac{k_8[Mad]\text{ [p55cdc}_\text{A}]}{J_8+\text{ [p55cdc}_\text{A}]}-k_6\text{ [p55cdc}_\text{A}],\\
&\frac{d\text{[Plk1]}}{dt}=k_9\text{[volume][CycB](1-[Plk1])}-k_{10}\text{[Plk1]},
\end{align}

\noindent where $k_i$ are the rate constants, the values being chosen in proportional to those in Tyson and Novak \cite{Novak2001, Novak2003} as given in Powathl et al. \cite{Powathil2012b}. The rate constants are scaled in such a way that it gives an average cell-cycle length of 20-25 hours under normal conditions. The parameter values of the cell-cycle model are scaled in such a way that each MCS step corresponds to 1 hour in real time and a normal cell has an average cell-cycle length of 25-35 hours while a cancer cells proliferate twice as faster.
For simulations involving tumour growth, the GGH target volume is incremented each MCS during the growth phases at a constant rate of 0.5 times the current cell volume. The cell-cycle model is incorporated into the CompuCell3D framework at each MCS using Bionetsolver and the SMBL ODE solver library. 

\subsection*{The tumour microenvironement}
Overall tumour progression and cell-cycle dynamics are critically dependent on the surrounding tissue microenvironment and in particular, the availability of oxygen. Here, the temporal and spatial evolution of the oxygen concentration distribution is modelled using the following partial differential equation \cite{Powathil2012b} and is incorporated into the CompuCell3D framework as a diffusive chemical field:

\begin{align}
&\frac{\partial K(x,t) }{\partial t}=\nabla.(D_K(x)\nabla K(x,t))+r(x) m(\Omega)-\phi K(x,t) \text{cell}(\Omega,t)
\label{oxygen_equation}
\end{align}

\noindent where $K(x,t)$ denotes the oxygen concentration at (pixel) position $x$ at time $t$, $D_K(x)$ is the diffusion coefficient and $\phi$ is the average rate of oxygen consumption by a cell with an area (in 2 dimensions) $\Omega$ at time $t$ ($\text{cell}(\Omega,t) =1$ if position $x\in \Omega$ is occupied by a cell at time $t$ and zero otherwise). Here, $m(\Omega)$ denotes the vessel cross section with an area $\Omega$ ($m(\Omega) =1$ for the presence of blood vessel at position $x \in \Omega$, and zero otherwise); thus the term $r(x)  m (\Omega)$ describes the production of oxygen at an average rate $r(x)$ (averaged over the vessel cross section area). This equation is solved using no-flux boundary conditions and an appropriate initial condition \cite{Powathil2012}. We have used the same parameters values as Powathil et al. \cite{Powathil2012b} and for the consistency we simulated the diffusion equation for the oxygen 1000 times in every 1 MCS to achieve the similar time scale of $0.001$ hr \cite{Powathil2012b}. We assume that hypoxia up-regulates the HIF 1 (Hypoxia Inducible Factors) pathway, which further results in changes in intracellular cell-cycle dynamics. When the average oxygen concentration at a specific cell location $\Omega$ falls below 10$\%$ (hypoxic cell), HIF-1 $\alpha$  is assumed to become active from an inactive phase, which further delays the cell-cycle dynamics (cf. Equation (1)).

\subsection*{Radiation response}
Following Powathil et al. \cite{Powathil2013}, the direct effects of radiation is incorporated into the multiscale model using the survival probability after the irradiation, given by:
\begin{align}
S(d)=\exp\left[\gamma \left(-\alpha.\text{OMF}.d-\beta (\text{OMF}.d)^2 \right) \right].
\label{SensitivityLQ}
\end{align}

\noindent where $d$ is the radiation dose and $\alpha$ and $\beta$ are sensitivity parameters, taken to be $\alpha=0.3~ \mathrm{Gy}^{-1}$ and $\beta=0.03~ \mathrm{Gy}^{-2}$ \cite{Powathil2012}. The effect of changing tissue oxygen levels on the radiation sensitivity is incorporated by using the concepts of an ``oxygen enhancement ratio" or ``oxygen modification factor" \cite{Powathil2012}, defined as

\begin{align}
\text{OMF}=\frac{\text{OER}(pO_2)}{\text{OER}_m}=\frac{1}{\text{OER}_m}\frac{\text{OER}_m.pO_2(x)+K_m}{pO_2(x)+K_m}
\label{OMF}
\end{align}
where $pO2(x)$ is the oxygen concentration at position $x$, \text{OER} is the ratio of the radiation doses needed for the same cell kill under anoxic and oxic conditions, $\text{OER}_m=3$ is the maximum ratio and $K_m = 3$ mm Hg is the pO2 at half the increase from 1 to $\text{OER}_m$ \cite{Powathil2012, Titz2008}. We have incorporated cell-cycle dependent varying sensitivity by an additional term $\gamma$ and it varies from 0 to 1, depending on the individual cell's position at the time of the irradiation. Here, we assumed that the cells in S-G2-M phase has maximum sensitivity with $\gamma=1$ while the cells in G1 phase  and the resting phase has relative sensitivities of $\gamma=0.5$ and  $\gamma=0.25$, respectively. We also accommodate cellular repair after the low-doses of irradiation and assume about 50\% of the DNA damages are likely to be repaired within few hours, increasing the survival chances of the cells and hence the final survival probability is written as:
\begin{eqnarray}\label{LQ_repair}
S^*(d)=\left\{
\begin{array}{lcl}
S\qquad \qquad \qquad \qquad \qquad \quad ~ d>5\\
\\
S+(1-S)\times 0.5 \qquad \qquad d\le 5.\\
\end{array}
\right. 
\end{eqnarray}

This survival probability is then used to calculate the survival chances of each cell when they are irradiated with the radiation rays. To study this survival chance of an individual cell, a random number is drawn for each cell at every time when they are irradiated and compared against the calculated survival probability. The irradiated cell survives if the random number is smaller than the survival probability and die otherwise.

Here, we also consider the effects of radiation on cell-cycle as irradiation results in a divisional delay, and, in particular, G2 phase delay/arrest in many cell lines \cite{Pawlik2004, Maity1994}. Experimental results show that irradiated cells in G2 phase may take up to 9 hours longer to complete the cell-cycle due to the activation of several intracellular repair mechanisms induced by the radiation \cite{Maity1994}. Radiation damage can also induce a cell-cycle delay  in G1 phase, mainly through the activation of p53 and p21 pathways \cite{Pawlik2004}. In the present model, we include this effect of irradiation induced delay by forcing the cells to stay in the same phase for an extra time duration of up to 9 hours \cite{Maity1994}. 

The indirect effects of radiation exposure is assumed to occur through the productions of radiation-induced bystander signals to which surviving irradiated cells and non-irradiated cells respond. The spatio-temporal evolution of these bystander signals are modelled using a reaction-diffusion equation, given by

 \begin{align}
&\frac{\partial B_s(x,t) }{\partial t}=D_{s}(x)\nabla^2 B_s(x,t)+r_{s} \text{cell}_{\text{Rad}}(\Omega,t)-\eta_{s}B_s(x,t)
\label{chemo_equation}
\end{align}
where $B_s(x,t)$ denote the strength or concentration of the signal at position $x$ and at time $t$, $D_s(x)$ is the diffusion coefficient of the signal (which is assumed to be constant), $r_s$ is the rate at which the signal is produced by a irradiated cell, $\text{cell}_{\text{Rad}}(\Omega,t)$ ($\text{cell}_{\text{Rad}}(\Omega,t)=1$ if position $x\in \Omega$ is occupied by a signal producing irradiated cell at time $t$ and zero otherwise) and $\eta_s$ is the decay rate of the signal. The bystander cells will then respond to these signals in multiple ways with various probabilities when the signal concentration is higher than certain assumed threshold.

\subsection*{Parameter Estimation}

 Most of the parameters related to the cell-cycle model and oxygen dynamics are chosen from previous mathematical and experimental papers. In the cell-cycle model, the parameter values are chosen from Tyson and Novak's papers \cite{Novak2001, Novak2003} but with time scales relevant to mammalian cell-cycle (Table \ref{table_parameters1}). Here, we have assumed a cell-cycle time of 20-30 $hrs$ under favourable conditions for a normal cell. The rest of the parameter values of the cell-cycle model can be found in Table \ref{table_parameters1}.

\begin{table}[ht] 
\caption{cell-cycle model parameters from \cite{Novak2001, Novak2003, Powathil2012b} (after rescaling to obtain a normal cell-cycle time of 20-25 hrs)} 
\centering 
{\footnotesize
\begin{tabular}{c c c c}
 \hline\hline
 &Component & Rate constants ($hr^{-1}$) & Dimensionless constants  \\
  [0.5ex]	
  \hline 
&[CycB]& $k_1=0.24$, $k'_2=0.24$, $k^{''}_2=9$, [p27/p21]=2.1& $[\text{CycB}]_{th}=0.1$\\ 
 &[Cdh1]&$k'_3=6$,$k^{''}_3=60$, $k_4=210$ &$J_3=0.04$,$J_4=0.04$\\
 & [p55cdc$_\text{T}$]& $k'_5=0.03$, $k^{''}_5=1.2$, $k_6=0.6$& $J_5=0.3$, $n=4$\\ 
 & [p55cdc$_\text{A}$]&$k_7=6$,$k_8=3$&$J_7=0.001$,$J_8=0.001$,[Mad]=1\\
 &[Plk1]& $k_9=0.6$, $k_{10}=0.12$& \\ 
  [1ex]
   \hline 
   \end{tabular} }
   \label{table_parameters1} 
   \end{table}

The oxygen dynamics are governed by a reaction diffusion equation, where the parameters are mainly chosen from the literature \cite{macklin}.  The oxygen diffusion length scale $L$ is often considered to be approximately equal to 100 $\mu$m and the value of the diffusion constant is taken as $2*10^{-5}$ cm$^2$/s \cite{Owen2004}. Using these and the relation $L=\sqrt{D/\phi}$, the mean oxygen uptake can be approximately estimated as 0.2 s$^{-1}$. The oxygen supply through the blood vessel is approximately taken as $8.2*10^{-3}$mols s$^{-1}$ \cite{Avneer2009}. 

\begin{figure}[h!]
  \begin{center}
      \includegraphics[scale=1]{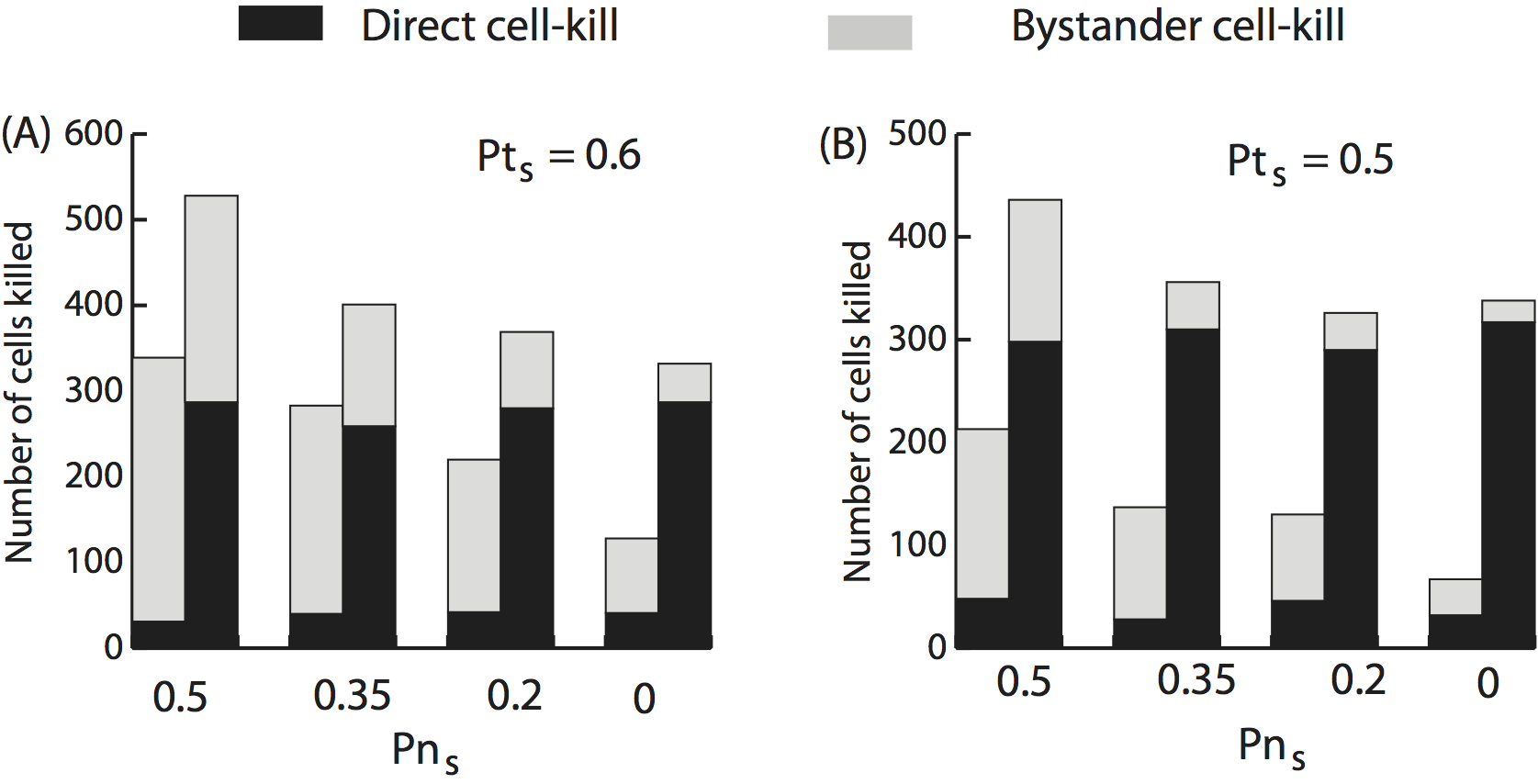}
      \end{center}
  \caption{Sensitivity analysis of the probabilities that determine the production of bystander signals. Figure shows the number of cells killed when tumour cells receive 0.25 Gy and 2 Gy dosage and normal cells receive a decreasing dosage (case 1) for various combination of probabilities}
  \label{Production}
\end{figure}

\begin{figure}[h!]
  \begin{center}
      \includegraphics[scale=1]{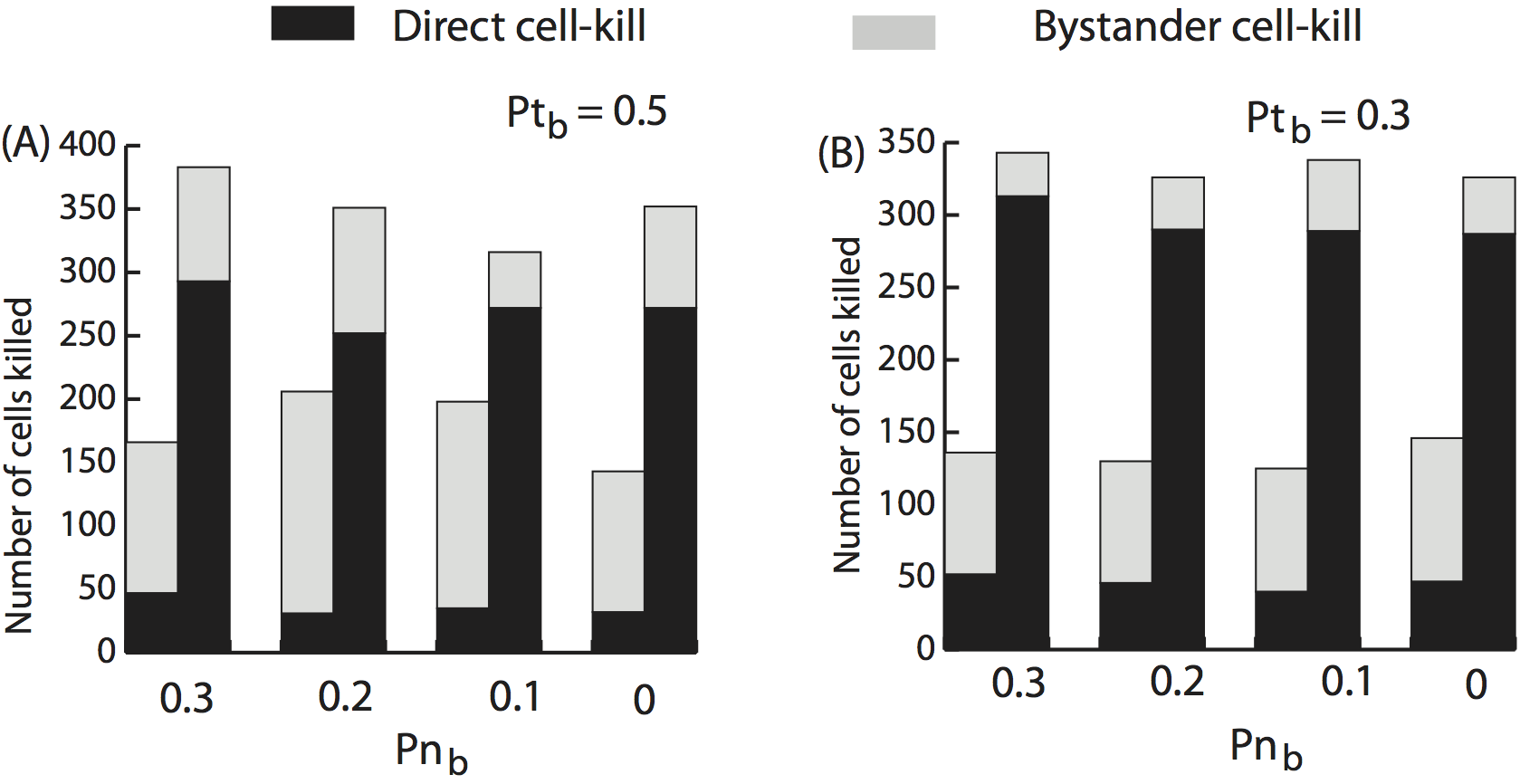}
      \end{center}
  \caption{Sensitivity analysis of the probabilities that determine the response of bystander cells to bystander signals. Figure shows the number of cells killed when tumour cells receive 0.25 Gy and 2 Gy dosage and normal cells receive a decreasing dosage (case 1) for various combination of probabilities}
  \label{Response}
\end{figure}
\begin{figure}[h!]
  \begin{center}
      \includegraphics[scale=1]{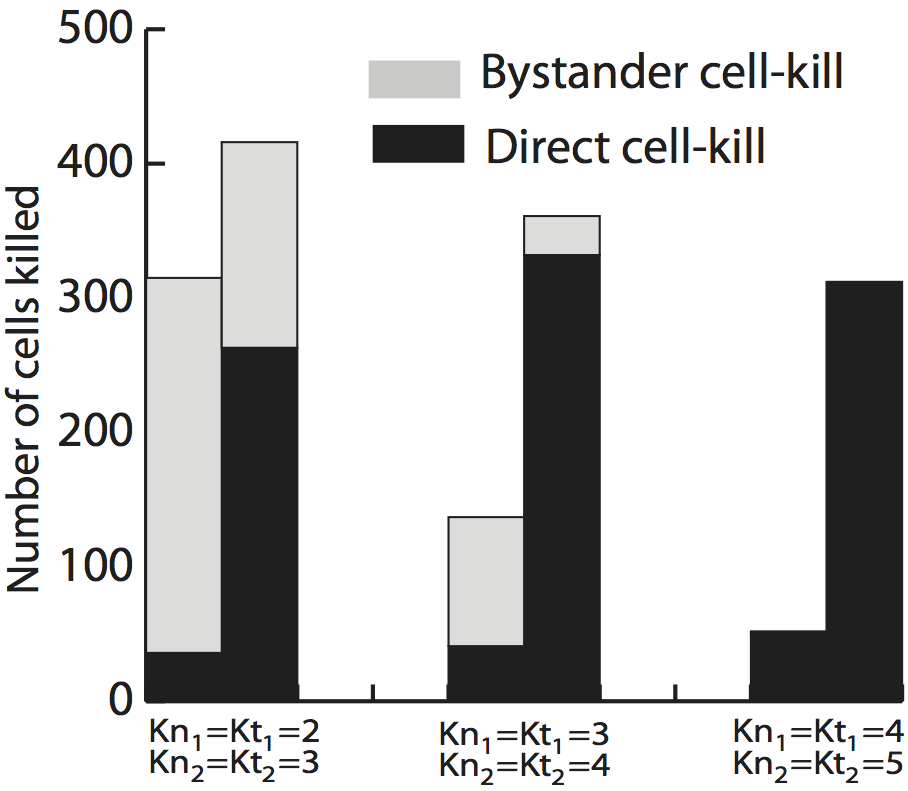}
      \end{center}
  \caption{Sensitivity analysis of the thresholds above which bystander cells respond to bystander signals. Figure shows the number of cells killed when tumour cells receive 0.25 Gy and 2 Gy dosage and normal cells receive a decreasing dosage (case 1) for three different threshold levels}
  \label{Threshold}
\end{figure}

\subsection*{Sensitivity Analysis}

In the present model, the dynamics of bystander signals and responses of bystander cells to these signals are analysed based on probabilistic approach as it is hard to determine the exact probabilities by which these bystander responses of normal or tumour cells occur \cite{Mothersill2006}. To study how these probabilities (signal production: $Pn_s$ and $Pt_s$ and bystander response: $Pn_b$ and $Pt_b$ in Figure 2 of main manuscript) affect on the direct and indirect radiobiological responses and its contributions to total cell-kill, a sensitivity analysis is carried out here by varying these probabilities and comparing the total cell-kill. Figure \ref{Production} shows the total cell-kill and the contributions from the direct and bystander cell-kills with respect to varying probabilities that determine the total number of bystander signal producing cells. Figure \ref{Production}.A shows the cell-kill when 60\% of the radiation exposed tumour cells and 50\%, 35\%, 20 \% and 0\% normal cells produce bystander signals. Here, the probabilities are chosen in such a way that $Pt_s >Pn_s$ as observed. The bar plots indicate that the contribution from the direct cell-kill estimated using the modified LQ model relatively same for all the combinations, while the bystander cell-kill increased with an increase in the probability of signal producing normal cells ($Pn_s$), as expected. A similar inference but with reduced bystander cell-kill can be deduced from the Figure \ref{Production}.B where the the probability for a irradiated tumour cells to produce bystander signal is 10\% lower. The effects of varying probabilities for a bystander tumour or normal cell to respond to the surrounding bystander signals is plotted in Figure \ref{Response}. The plots show that the minor variation in the probabilities has no significant effects on the final cell-kill and as seen in the above case, the direct cell-kill remains mostly the same. Please note that the responses of the bystander cells are also affected by the concentration of the bystander signals and as one expect, a lower threshold ($Kn_1,~Kn_2,~ Kt_1$ and $Kt_2$) can increase the number of bystander cells that are susceptible for bystander responses (as shown in Figure \ref{Threshold}). 

\begin{figure}[h!]
  \begin{center}
      \includegraphics[scale=.6]{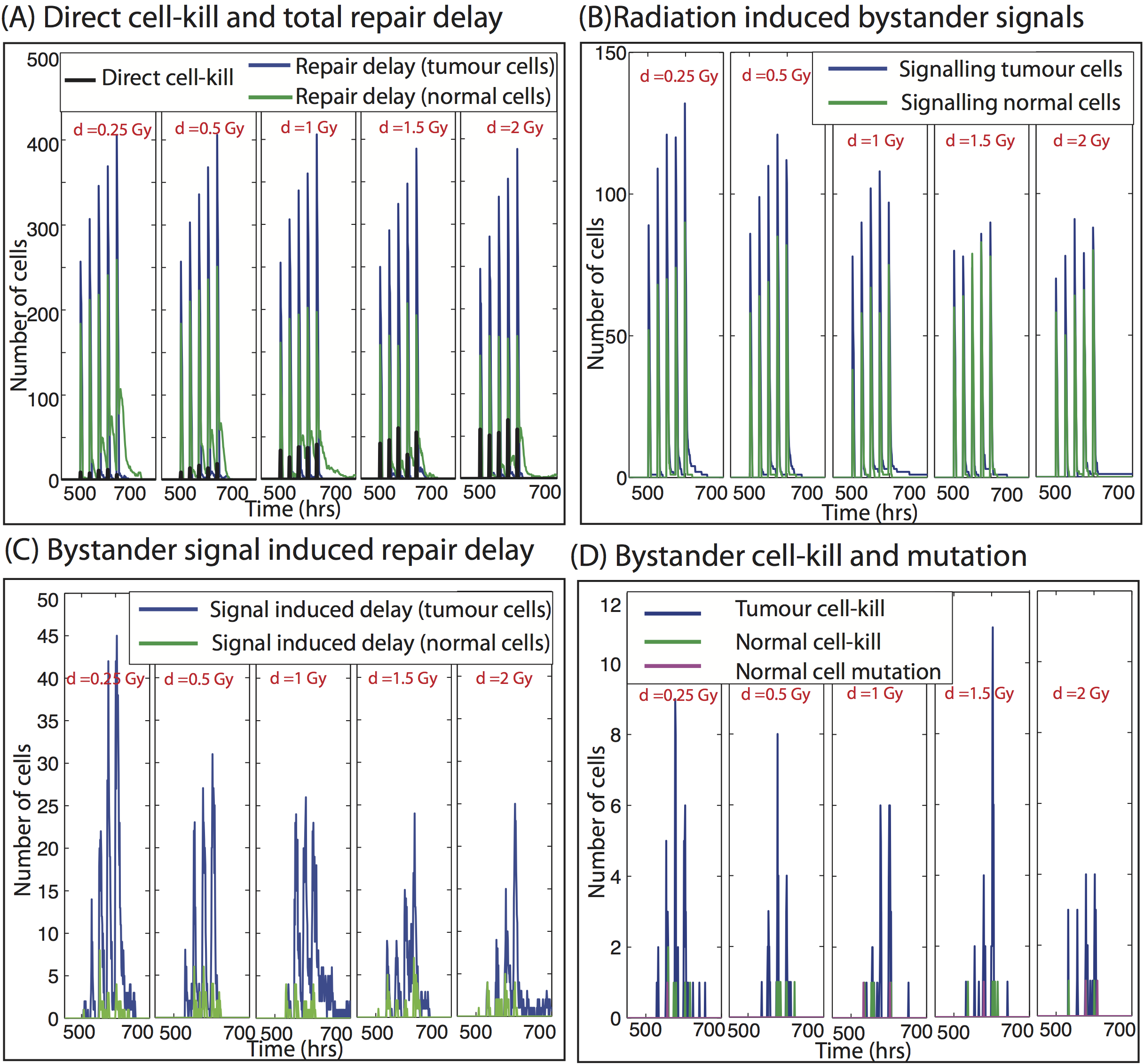}
      \end{center}
  \caption{Plots showing the direct and indirect effects of radiation when tumour cells are exposed to full given radiation dose while the surrounding normal cells receive gradient of doses. (A) Plots show various direct effects of irradiation for multiple doses, (B) Plots show the number of cells producing bystander signals, (C) Plots show number of cells with bystander signal induced repair-delay and (D) Plots show the effects of bystander signals on normal and tumour cells}
  \label{gradient2}
\end{figure}

\begin{figure}[h!]
  \begin{center}
      \includegraphics[scale=.6]{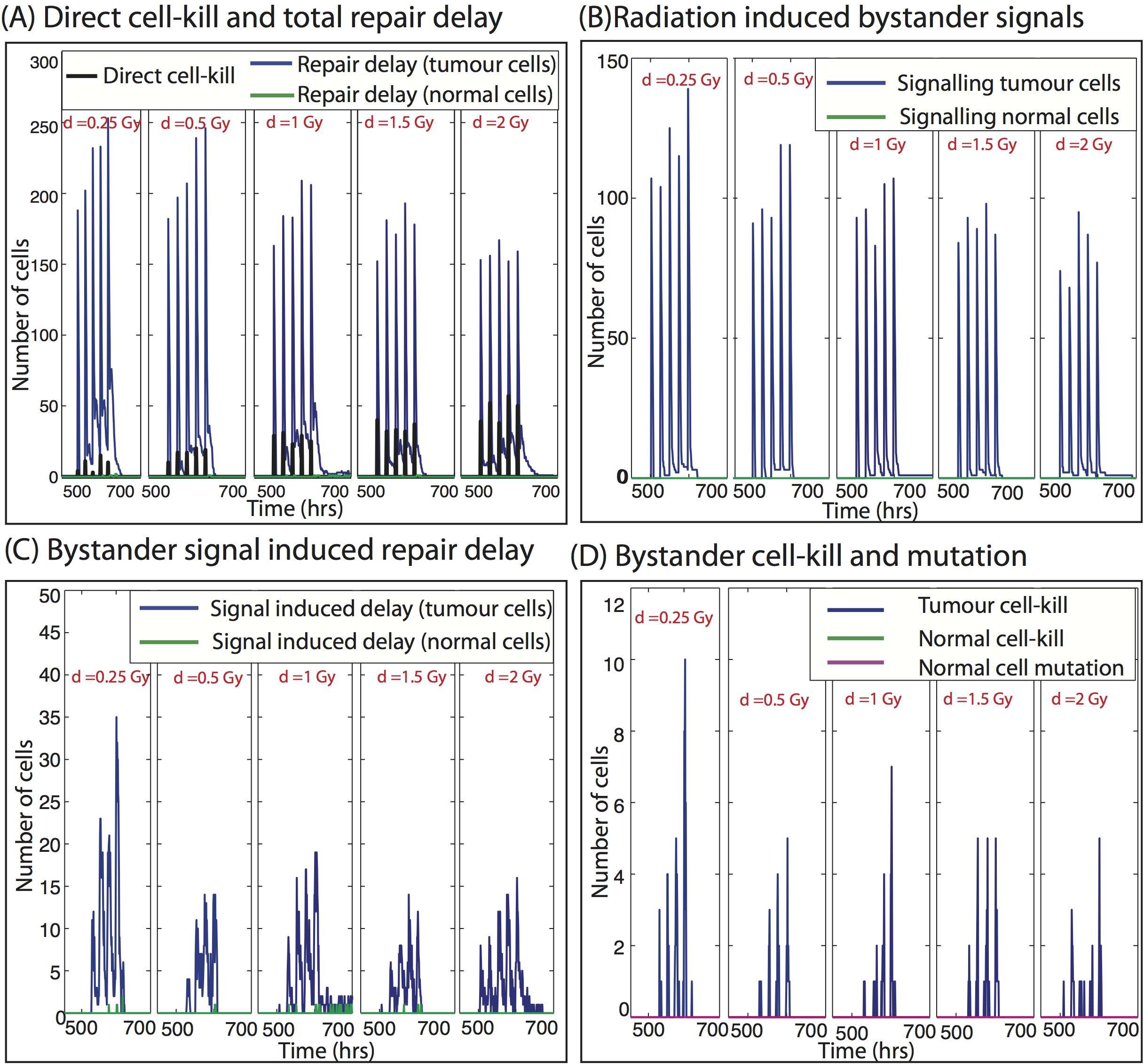}
      \end{center}
  \caption{Plots showing the direct and indirect effects of radiation when tumour cells are exposed to full given radiation dose while the surrounding normal cells are spared completely. (A) Plots show various direct effects of irradiation for multiple doses, (B) Plots show the number of cells producing bystander signals, (C) Plots show number of cells with bystander signal induced repair-delay and (D) Plots show the effects of bystander signals on normal and tumour cells}
  \label{tumour}
\end{figure}

\begin{figure}[h!]
  \begin{center}
      \includegraphics[scale=.6]{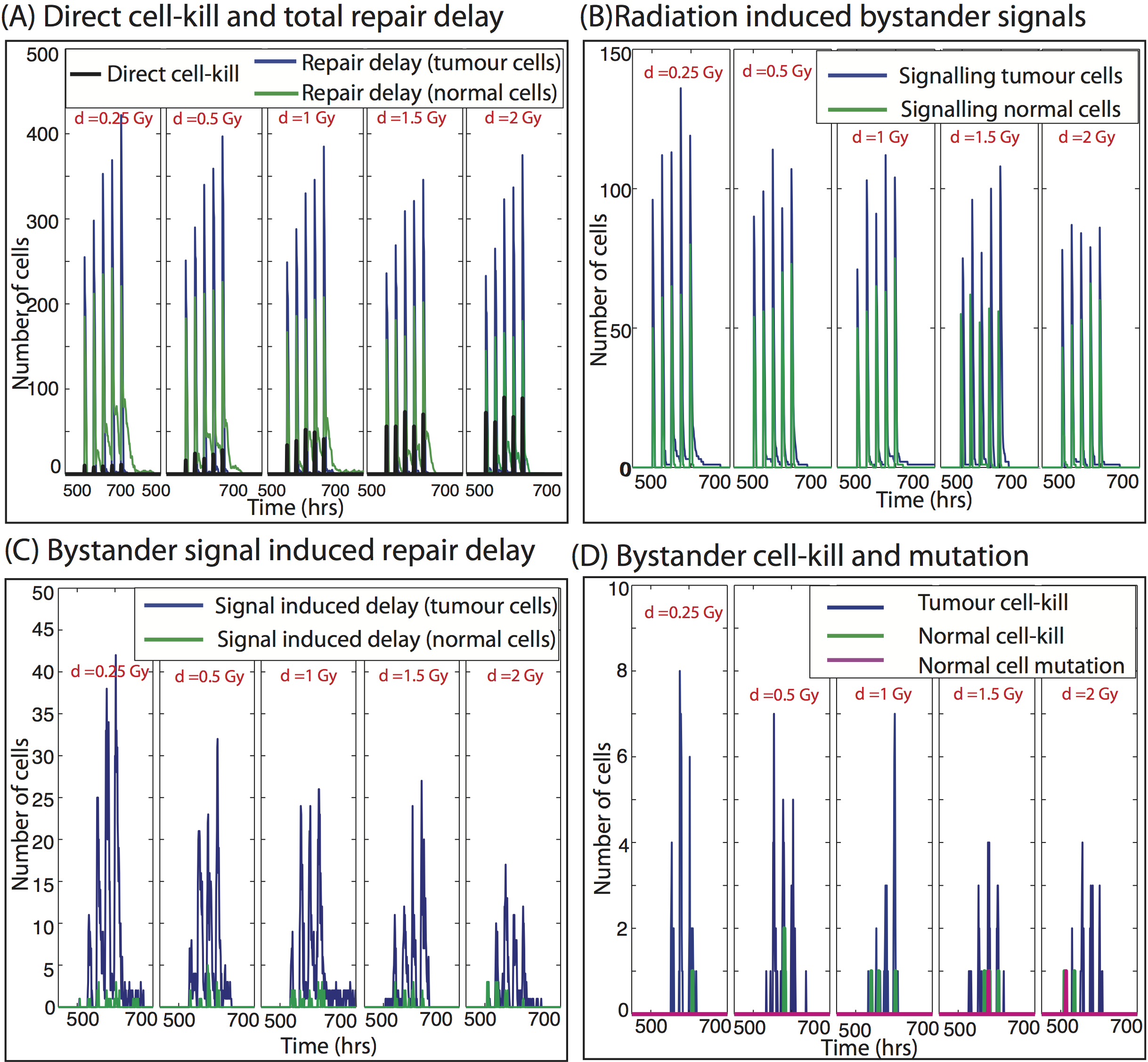}
      \end{center}
  \caption{Plots showing the direct and indirect effects of radiation when tumour and normal cells are exposed to full given radiation dose. (A) Plots show various direct effects of irradiation for multiple doses, (B) Plots show the number of cells producing bystander signals, (C) Plots show number of cells with bystander signal induced repair-delay and (D) Plots show the effects of bystander signals on normal and tumour cells}
  \label{fulldose}
\end{figure}

\subsection*{Direct and Bystander Effects}
\subsubsection*{Case 1: Targeting normal and tumour cells with varying dosage}
Figure \ref{gradient2}.A shows the radiation-induced cell-kill and the number of cells under repair delay for various doses per fraction. The plots show the radiation-induced cell-kill (both tumour and normal) and the number of cells that undergo radiation or bystander signal induced repair delay in their respective cell-cycle phases. We have assumed that the radiation can induce a cell-cycle delay, forcing cells to stay in the same cell-cycle phase for an extra time duration of up to 9 hours \cite{Powathil2013}. Additionally, depending on the intensity and the probability, bystander signals may also induce a cell-cycle delay of upto 6 hours \cite{Hu2006}. The plots show an increased cell-kill when the host-tumour system is irradiated with doses greater than 0.5 Gy and at low dose rates, more normal cells are under radiation-induced repair delay. Figure \ref{gradient2}.B shows the total number of cells that survive the cell-kill due to direct irradiation but undergo cell-cycle delay to repair the DNA damage and produce further radiation-induced bystander signals. At low doses there is a higher number of signalling tumour cells and normal cells, whilst the number of signalling tumour cells is lower at high doses due to increased cell-kill. In addition to these signalling cells, the cells that undergo radiation-induced cell-death (but are still-alive) are also assumed to produce bystander signals. We assume that the dead cells do not emit bystander signals. The bystander responses to the radiation-induced bystander signals are given in Figures \ref{gradient2}.C and \ref{gradient2}.D. As expected, the number of cells undergoing cell repair delay is higher for low doses as more cells are exposed to the moderate intensity bystander signals. At high dose rates, the increased cell-kill results in more localised sources of the diffusing bystander signal and although the number of cells under repair-delay is low during radiation, more cells are being exposed to moderate signal intensities after radiation.

\subsubsection*{Case 2: Targeting tumour cells}

Figures \ref{tumour}.A and \ref{tumour}.B show the direct cell-kill as well as the total number of cells under repair delay and the total number of signalling cells. As normal cells are spared from radiation exposure, none of the normal cells undergo radiation-induced cell-cycle delay and although the number of tumour cells under delay is higher for low-dose radiation, it is significantly lower when high doses per fraction are given. Consequently, in this scenario, only the tumour cells produce the radiation-induced bystander signals and thus will have less effect on the surrounding normal cells. The effects of radiation-induced bystander signals produced by irradiated tumour cells are given in the Figures \ref{tumour}.C and \ref{tumour}.D. The plots show that while the bystander signals induce cell-cycle delay and cell-kill within the tumour cells, they have no major effects on normal cells as cells are assumed to respond to the bystander signals only when signal concentration is above a threshold  level. This is a therapeutic ideal - reducing the normal tissue damage while still maintaining the tumour control, and may not be achievable using the current clinical delivery methods.

\subsubsection*{Case 3: Targeting normal and tumour cells with full dose}

Figure \ref{fulldose}.A shows an increased cell-kill due to the direct effects and a similar distribution of total cells that are under the repair delay. Figure \ref{fulldose}.B shows that the number of signalling normal cells is lower compared to case 1, since all the cells receive high dose and are more likely to be killed directly. The bystander responses plotted in Figures \ref{fulldose}.C and \ref{fulldose}.D show that the bystander responses are similar to that of case 1 at low doses whilst fewer cells responded to the bystander signals at high doses due to a weaker bystander signals concentration. Here in case 3, although the full dose delivery to the host-tumour system increases the overall cell-kill, it comes in an expense of normal cell-kill and radiation damage, which should be avoided in ideal scenario.

\small